\documentclass[11pt, a4paper]{article}
\pdfoutput=1 

\usepackage[height=8.85in,width=6.75in]{geometry}

\usepackage{graphicx,rotating}     
\usepackage[bookmarksopen,colorlinks=true,linkcolor=steelblue,
citecolor=orangered,urlcolor=darkred,linktoc=all]{hyperref}

\usepackage{amsmath,amssymb}
\usepackage{slashed}
\usepackage{bm}
\usepackage{bbm}
\usepackage{cite}
\usepackage{comment}
\usepackage[utf8]{inputenc}
\usepackage{accents}
\usepackage[footnotesize]{caption}
\usepackage{subcaption}
\usepackage{cancel}

\newcommand{\abs}[1]{\left\vert {#1} \right\vert}

\newcommand{\mcp}{{\chi}}
\newcommand{\mcpb}{{\bar{\chi}}}
\newcommand{\abb}[1]{\left\vert \boldsymbol #1 \right\vert}
\newcommand{\millicharge}{{\varepsilon}}

\usepackage{xcolor}
\definecolor{darkred}{rgb}{0.7, 0., 0.}
\definecolor{orangered}{rgb}{1,0.27,0.}
\definecolor{steelblue}{rgb}{0.275,0.51, 0.706}
\definecolor{forestgreen}{rgb}{0.13,0.55,0.13}

\usepackage{tikz}
\usepackage{tikz-feynman}
\tikzfeynmanset{compat=1.0.0}
\pgfdeclarelayer{bg}    
\pgfsetlayers{bg,main}  

\begin{document}


\begin{center}


\vskip 0.5in

{\Large \bfseries 
Constraints on millicharged particles from nuclear $\gamma$-decays
} \\
\vskip .8in

{Ting Gao$^{1,a}$, \let\thefootnote\relax\footnote{$^a$gao00212@umn.edu}
Maxim Pospelov$^{1,2,b}$ \footnote{$^b$pospelov@umn.edu}}

\vskip .3in
\begin{tabular}{ll}
$^1$& \!\!\!\!\!\emph{School of Physics and Astronomy, University of Minnesota, Minneapolis, MN 55455, USA}\\
$^2$& \!\!\!\!\!\emph{William I. Fine Theoretical Physics Institute, School of Physics and Astronomy,}\\[-.3em]
& \!\!\!\!\!\emph{University of Minnesota, Minneapolis, MN 55455, USA}
\end{tabular}

\end{center}
\vskip .6in

\begin{abstract}
\noindent
We consider nuclear gamma decays and $\gamma$-emitting reactions that can be an efficient source of hypothetical millicharged particles ($\chi$). In particular, we revisit the production of millicharged particles in nuclear reactor environment, pointing out that $\gamma$ cascades from $^{239}$U is an overlooked yet a powerful source of  $\chi\bar\chi$ pairs. This leads to an increased flux compared to previous studies. We then apply new estimates of the flux to derive novel limits on the value of millicharge, $\varepsilon = Q_\chi/e$, from the electron recoil searched for in a variety of experiments placed in proximity to the reactor cores. The derived limits on $\varepsilon$ are the strongest in the interval of masses $\sim 0.7-2 $\,MeV. We also derive the MCP flux from the Sun  and point out potential sensitivity of the low-threshold dark matter search experiments. 
\end{abstract}

\section{Introduction}

While Standard Model (SM) of particles and fields has been extremely successful in describing observable world, it cannot be a ``final theory". Notable exceptions include neutrino physics (mixing and oscillations), as well as cold dark matter (DM) that must be associated with new degrees of freedom. Generalization of existing ideas about neutrino masses, as well as particle dark matter leads to the notion of dark sectors (DS). DS constructions are not tied to any specific energy scale, and can be indeed much lighter than the scale of weak interactions. Light DS, by necessity, would have to have states with rather small couplings to the SM states, and their search would typically require high-intensity experiments \cite{Lanfranchi:2020crw,Beacham:2019nyx}. In light of this, all possible connections (or portals) between light DS and SM have been described, classified and searched for in ever increasing number of high-intensity experiments. One unique portal to DS is via the so-called millicharged particles (MCP).

In this paper, we will revisit some aspects of searches for the MCP $\chi$ that we will assume to be a fermion state $\chi$ with an electric charge much smaller than the proton charge, $ Q_\chi \equiv \varepsilon e \ll e$. We note that while such construction may look totally artificial, there are theoretically natural ways of ascribing the smallness of $\varepsilon$ to a small amount of mixing between regular and ``dark" photons \cite{Holdom:1985ag}. There has been a steady interest in phenomenology of  millicharged particles, not least because they are capable of inducing a variety of interesting phenomena. MCPs can bring change to the early Universe cosmology \cite{Davidson:2000hf,Vogel:2013raa,Dolgov:2013una}; they can be produced and accelerated by supernova \cite{Chang:2018rso,Dunsky:2018mqs,Fiorillo:2024upk}; their population on Earth can be enhanced due to possibly small collision lengths within Earth's material \cite{Pospelov:2020ktu,Berlin:2023zpn}. 
Several years ago, MCP dark matter was invoked as a possible explanation of the abnormal hydrogen spin temperature signal observed by the EDGES collaboration \cite{Bowman:2018yin} that consists in stronger-than-expected 21\,cm absorption feature at redshifts of $z\sim O(15-20)$. An explanation of such phenomenon may reside in late re-coupling of ordinary baryonic matter and dark matter, and that can naturally occur if DM has a small millicharge \cite{Barkana:2018qrx,Kovetz:2018zan,Liu:2019knx}. These scenarios, in particular, prefer relatively light, sub-100-MeV MCP particles with $\varepsilon\sim O(10^{-5})$.

All existing MCP limits and proposals for future searches can be subdivided into two categories: those that exploit production and detection of MCPs and those that search/constrain DM-style MCP relics. 
Any signal in the first method would typically scale as the fourth power of $\sim \varepsilon$, while the second class of searches is $\sim \varepsilon^2$, but is more model-dependent\footnote{An interesting class of its own is the search of MCPs via missing energy/missing momentum, that also has $\varepsilon^2$ type of scaling \cite{Berlin:2018bsc}}. We will concentrate on the first method that involves the production and detection of the MCPs, and limit our study to relatively light, sub-10-MeV particles. 
Specifically, the MCPs were searched for in the electron beam dump experiments \cite{Prinz:1998ua} that provide the most direct limits in the sub-100-MeV range. Similar style limits can be derived from the past proton beam dump/neutrino experiments \cite{Magill:2018tbb}, and new dedicated experiments placed along the line of sight of various particle beams \cite{Kelly:2018brz,Harnik:2019zee,ArgoNeuT:2019ckq,Kling:2022ykt,SENSEI:2023gie,Tsai:2024wdh,Essig:2024dpa}. Moreover, cosmic-ray-initiated production of MCPs can be used to derive equally stringent limits on $\varepsilon$ \cite{Plestid:2020kdm,Harnik:2020ugb,Du:2022hms,Wu:2024iqm}.

If masses of DS particles are light, below 10\,MeV, one must anticipate that the strongest direct constraints may come from the particle physics experiments in direct proximity of nuclear reactors. This is because commercial nuclear reactors can be viewed as very powerful beam dump experiments, where nuclear fission generates copious amounts of $\bar\nu,\gamma,n$ radiation with $O({\rm MeV})$ energy spectra. Recent years have seen the proliferation of low-threshold detectors installed in $O(30\,{\rm m})$ distance from cores of nuclear reactors with the main goal of detecting coherent neutrino-nucleus recoil. Low threshold achieved in these experiments also help to set stringent constraints on MCPs, as typical cross section for detection scales inversely proportional with the kinetic energy of the outgoing electron, $E_{\rm recoil}^{-1}$. Several collaborations have placed limits on MCPs using such devices \cite{TEXONO:2018nir,CONNIE:2024off}. Previous searches of the neutrino magnetic moments as well as actual measurements of the $\bar\nu e$ scattering can also be used to be recast as limits on the MCPs. The intense photon flux in nuclear reactors also provide an efficient environment to probe other DS particles, such as axion-like particles, as has been done in \cite{TEXONO:2006spf,Dent:2019ueq,AristizabalSierra:2020rom}.

Several experimental studies \cite{TEXONO:2018nir,CONNIE:2024off} have recognized the potential of nuclear reactors and the enormous photon flux that they generate as an important source of MCPs. They derived stringent limits on such particles already. Our present study is motivated by the observation that existing treatments for MCPs \cite{TEXONO:2018nir,CONNIE:2024off} seem to have rather limited reach in terms of $m_\chi$ stemming from the fact that only $e+\gamma \to e+ \bar\chi + \chi$ reactions are included as the source of MCPs. Yet it is clear that {\em any} nuclear process resulting in emission of a gamma quantum of energy $\omega$ can in principle emit a charged pair \cite{Kroll:1955zu}. We adopt this production mechanism for nuclear reactors, following recent treatment of pair emission in Ref. \cite{Pitrou:2019pqh}. We find that the stringent limits can be extended to higher masses.

While nuclear reactors provide the largest fluxes of MCPs for the closely located detectors, we note that there is also an ambient flux of MCP created by natural radioactivity (mostly caused by the radioactive isotopes of K, U and Th). While fluxes of MCP created this way are going to be much lower than those in proximity of reactor cores, the detection methods that can be employed could involve neutrino detectors, and large self-shielding dark matter detectors (such as those based on xenon dual-TPC technology). These are large deep-underground detectors operating in a maximally background-free environment, capable of achieving much lower counting rates than detectors placed in reactors' proximity. Similarly, the MCPs can be created as a byproduct of nuclear reactions in the Sun. Here, however, the value of $\epsilon$ must be much smaller to avoid the slow-down and thermalization of MCPs by the solar matter. 

Along with the new estimates of fluxes from MCPs, we also calculate the production of massive dark photons. If the mass is above $2m_e$, the dark photons are unstable against the decays to $e^+e^-$. If kinematically allowed, they can be produced in nuclear reactors, and possibly detected by {\em e.g.} experiments studying neutrino-electron scattering.

The rest of our paper is organized as follows: Section 2 of this work describes the production of MCPs accompanying nuclear decays. Section 3 treats different experiments and derives corresponding limits on $m_\chi-\varepsilon$ plane. Section 4 estimates the fluxes and detection rates of MCPs from natural radioactivity and solar production. Section 5 addresses limits on massive dark photons.  We reach our conclusions in section 6.

\section{Pair-production from nuclear de-excitation}

Any process that results in the emission of an on-shell $\gamma$ of frequency $\omega$ may also produce a pair of MCP particles with $2m_\chi< \omega$. For nuclear-initiated emission, the recoil is negligible, and therefore the entire energy of a photon can go into the MCP pair, so that the kinematic limit for $m_\chi$ is close to $\omega/2$. For photons emitted from the electron lines, in contrast, the kinematic limit is lower due to the electron recoil. In previous analyses \cite{TEXONO:2018nir,CONNIE:2024off}, the main source of the MCP was assumed to originate from $e+\gamma\to e+ \chi + \bar \chi $ reaction. This results in a sharp drop of the $\chi$ flux for $m_\chi > 0.5$\,MeV. We note, however, that the spectrum of photons emitted in nuclear fission is extended into a several MeV range, and therefore higher $m_\chi$ than those considered in Refs. \cite{TEXONO:2018nir,CONNIE:2024off} can be emitted. The photons can be emitted in the de-excitations of the fission products, and in the process of the subsequent capture of free neutrons. As $\sim$60\% of the neutrons produced in fission undergo mostly $(n,\gamma)$ reactions on protons or on heavier elements, there is a plenty of MCP sources that should be added to the production mechanisms discussed in \cite{TEXONO:2018nir,CONNIE:2024off}. 

In this section, we calculate the rate of $\mcp\mcpb$ pair production from neutron capture in the core region of the nuclear reactor. Each fission in the reactor releases an average of $200\mathrm{MeV}$ energy \cite{talou2023nuclear} and $2.5$ neutrons \cite{TEXONO:2005fmk}, which corresponds to  the neutron yield per second as
\begin{equation}
    N_n(\mathrm{s}^{-1})=7.8\times10^{19}\times \mathrm{Power(GW)}.
\end{equation}

Following their emission, neutrons undergo relatively rapid thermalization before either contributing to the next fission cycle or being captured by $^{238}\mathrm{U}$, by protons in natural-water reactors, or by other elements.
A large fraction of the neutron capture events leads to the emission of $\gamma$-rays, and, in the presence of millicharged particles, off-shell $\gamma^*$ which decays into a MCP pair. Following the procedure in \cite{Pitrou:2019pqh}, we first calculate the cross section ratio between these two processes
\begin{equation}
    r:=\frac{\sigma\left(n+N\to N'+\mcp+\mcpb\right)}{\sigma\left(n+N\to N'+\gamma\right)},
\end{equation}
and then use the well-studied $\gamma$-ray emission branching ratios to calculate the MCP pair production. The key assumption here is that the nuclear matrix elements for photon emission do not change in a significant way when the photon is taken off-shell. This is ultimately related to the small size of atomic nuclei compared to wavelengths of emitted photons and MCPs. 

\begin{figure}[t]
    \centering
    \begin{subfigure}[b]{0.49\textwidth}
    \includegraphics[width=\textwidth]{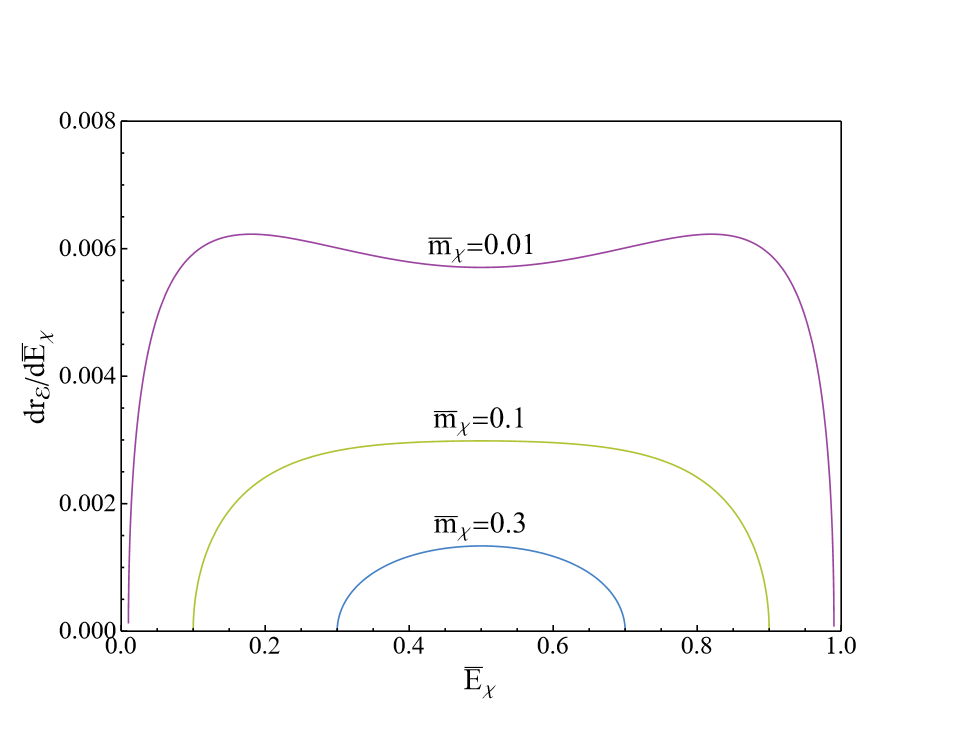}
    \caption{}
    \label{fig:rE}
    \end{subfigure}
    \begin{subfigure}[b]{0.49\textwidth}
    \includegraphics[width=\textwidth]{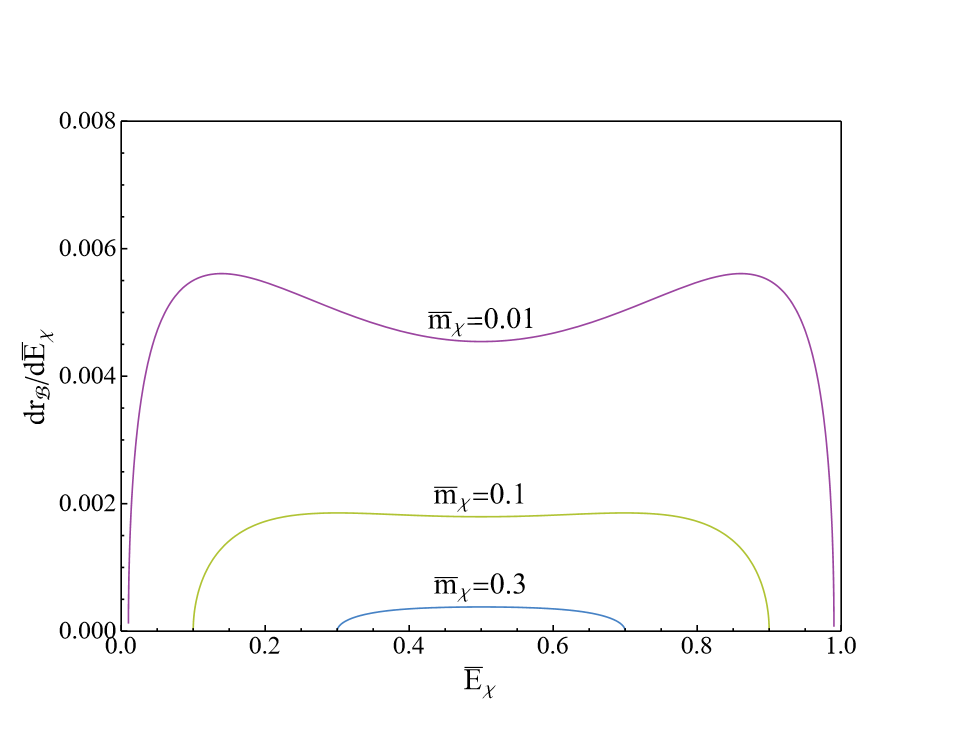}
    \caption{}
    \label{fig:rB}
    \end{subfigure}
    \caption{Differential ratios for E1 (\ref{fig:rE}) and M1(\ref{fig:rB}) transitions with $\millicharge=1$ for different values of $m_\mcp/\omega$. Quantities with a bar on top are normalized by $\omega$ for universality: $\bar{m}_\mcp=m_\mcp/\omega$, $\bar{E}_\mcp=E_\mcp/\omega$. Notice that when $m_\chi \ll \omega/2$, the E1 and M1 yields of MCP are similar, while for heavier MCP (relative to $\omega/2$), the M1 yields are more suppressed, as expected.}
    \label{fig:differential ratios}
\end{figure}

After a multipole expansion of the nuclear matrix element, the differential ratios for electric dipole (E1) and magnetic dipole (M1) transitions are given as
\begin{align}
    &\frac{dr_{\mathcal{E}(\mathcal{B})}}{dE_\mcp dE_{N'}}=\frac{M}{8\pi^2\omega}\mathcal{R}_{\mathcal{E}(\mathcal{B})},\label{differentialratio}\\
    &\mathcal{R}_{\mathcal{E}}=\frac{4\pi\alpha\millicharge^2}{\omega^2}\frac{1}{k^4}\left[\left(E_\mcp+E_\mcpb\right)^2\left(4m_\mcp^2+2k^2\right)-4k^2E_\mcp E_\mcpb+k^4\right],\\
    &\mathcal{R}_{\mathcal{B}}=\frac{4\pi\alpha\millicharge^2}{\omega^2}\frac{1}{k^4}\left\{4m_\mcp^2\left[\left(E_\mcp+E_\mcpb\right)^2-k^2\right]-k^4+2k^2\left(E_\mcp^2+E_\mcpb^2\right)\right\}.
\end{align}
Where $M$ is the mass of the excited state formed by neutron capture, $\omega=M-M_{N'}$ is the available energy, $k=p_1+p_2$ is the total momentum of the millicharged particle pair, and $\mathcal{R}_{\mathcal{E}(\mathcal{B})}$ is the ratio between the angular-averaged pair production and photon emission amplitudes for E1(M1) transition. We assume recoil-less processes, {\em i.e.} taking the leading order expansion in $1/M$ in the above and in the following discussion. The kinematical limits of $E_{N'}$ are determined by having $\mcp$ and $\mcpb$ moving either parallel or antiparallel to each other. After integrating over $dE_{N'}$, we get the differential ratios as functions of $E_\mcp$, which is convenient for the discussion of detection rates:
\begin{align}
    \frac{dr_\mathcal{E}}{dE_\mcp}=&\frac{\alpha\millicharge^2}{2\pi\omega^3}\left[2\abb{p_\mcp}\abb{p_\mcpb} +\frac{2m_\mcp^2 \omega^2 \abb{p_\mcp}\abb{p_\mcpb}}{\left(E_\mcp E_\mcpb+m_\mcp^2+\abb{p_\mcp}\abb{p_\mcpb}\right)\left(E_\mcp E_\mcpb+m_\mcp^2-\abb{p_\mcp}\abb{p_\mcpb}\right)}\right.\\
    &\nonumber\qquad\qquad\left.+\left(E_\mcp^2+E_\mcpb^2\right)\ln\frac{E_\mcp E_\mcpb+m_\mcp^2+\abb{p_\mcp}\abb{p_\mcpb}}{E_\mcp E_\mcpb+m_\mcp^2-\abb{p_\mcp}\abb{p_\mcpb}}\right],\\
    \frac{dr_\mathcal{B}}{dE_\mcp}=&\frac{\alpha\millicharge^2}{2\pi\omega^3}\left[-2\abb{p_\mcp}\abb{p_\mcpb} +\frac{2m_\mcp^2 \omega^2 \abb{p_\mcp}\abb{p_\mcpb}}{\left(E_\mcp E_\mcpb+m_\mcp^2+\abb{p_\mcp}\abb{p_\mcpb}\right)\left(E_\mcp E_\mcpb+m_\mcp^2-\abb{p_\mcp}\abb{p_\mcpb}\right)}\right.\\
    &\nonumber\qquad\qquad\left.+\left(E_\mcp^2+E_\mcpb^2-2m_\mcp^2\right)\ln\frac{E_\mcp E_\mcpb+m_\mcp^2+\abb{p_\mcp}\abb{p_\mcpb}}{E_\mcp E_\mcpb+m_\mcp^2-\abb{p_\mcp}\abb{p_\mcpb}}\right],
\end{align}
where
\begin{equation}
    E_\mcpb=\omega-E_\mcp,\qquad \abb{p_\mcp}=\sqrt{E_\mcp^2-m_\mcp^2},\qquad \abb{p_\mcpb}=\sqrt{(\omega-E_\mcp)^2-m_\mcp^2}.
\end{equation}
These results are plotted in Fig.\ref{fig:differential ratios}.

The type of neutron capture interactions along with their yield per fission are listed in \cite{TEXONO:2005fmk}\footnote{Although the study \cite{TEXONO:2005fmk} simulates the neutron capture yields $Y_n$ in specific environment of the reactor used by the TEXONO collaboration, we consider it to be a plausible estimate of a generic $Y_n$ for any natural water reactor. }, and the emission lines resulting from neutron capture are documented in the ENSDF database\cite{ENSDF}. The two most important reactions we consider is $p(n,\gamma){\rm ^2H}$ and ${\rm ^{238}U}(n,\gamma){\rm ^{239}U}$. With the goal of producing $\sim \mathrm{MeV}$ MCP pairs, we include the $3.297\mathrm{MeV}$ and $4.060\mathrm{MeV}$ E1 transitions from the decay of the $^{239}\mathrm{U}$ state at the threshold of $^{238}\mathrm{U}+n$ and the $2.223\mathrm{MeV}$ M1 transition from $^2\mathrm{H}$. The number of $\gamma$ emission events per second for each emission line is given as
\begin{equation}
    N_\gamma^{(N,\omega)}=N_n\times \frac{Y_n^{(N)}}{\sum Y_n}\times \frac{I_\gamma^{(N, \omega)}}{100},
\end{equation}
where $Y_n^{(N)}$ is the neutron capture yield on isotope $N$ per fission, and $I_\gamma^{(N, \omega)}$ is the number of $\gamma$ with energy $\omega$ emitted per 100 neutron captures on the isotope $N$.

The differential flux of millicharged particles is then found by
\begin{equation}
    \frac{d\phi_\mcp}{dE_\mcp}=\frac{2}{4\pi D^2}\sum_{N,\omega} \frac{dr_{\mathcal{E}(\mathcal{B})}}{dE_\mcp}N_\gamma^{(N,\omega)},
\end{equation}
where the factor of 2 comes from counting both $\mcp$ and $\mcpb$, and $D$ is the distance from the reactor core to the detector. The summation runs over the aforementioned transitions, and for both transitions of $^{239}\mathrm{U}$ we use $r_\mathcal{E}$, while for the $^2\mathrm{H}$ transition  we take $r_\mathcal{B}$.

\section{Millicharged particle detection via atomic ionization}
\begin{figure}[t]
    \centering
    \begin{subfigure}[b]{0.49\textwidth}
    \includegraphics[width=\textwidth]{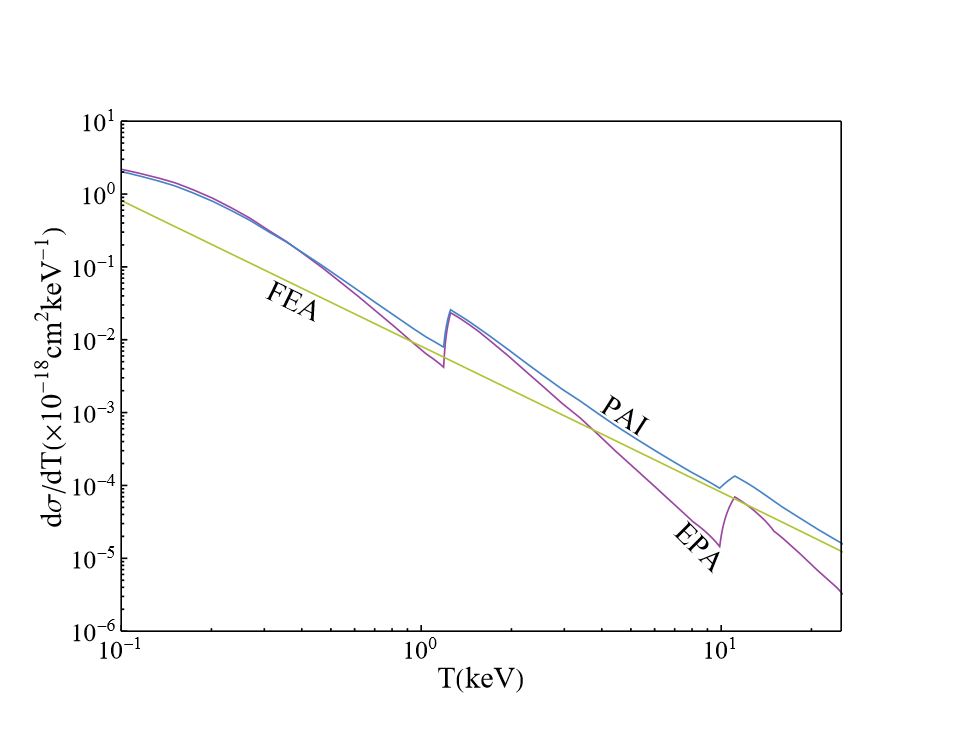}
    \caption{}
    \label{fig:sigma1}
    \end{subfigure}
    \begin{subfigure}[b]{0.49\textwidth}
    \includegraphics[width=\textwidth]{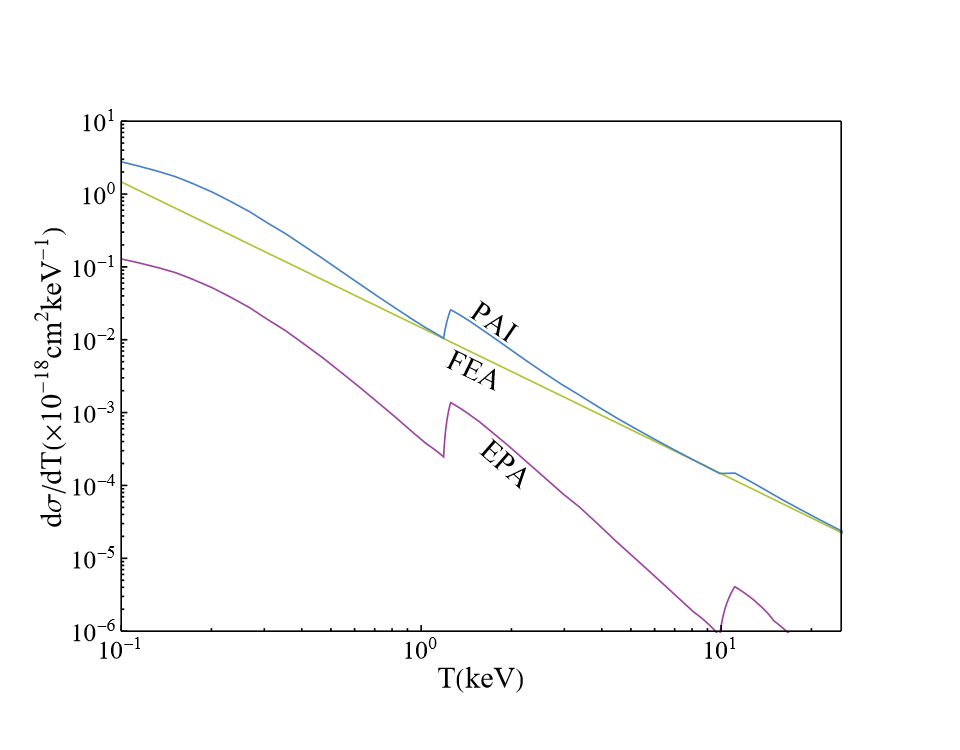}
    \caption{}
    \label{fig:sigma2}
    \end{subfigure}
    \caption{Atomic ionization differential cross sections for germanium crystal under Photo Absorption Ionization model (PAI), free electron approximation (FEA), and equivalent photon approximation (EPA) for $m_\mcp=1\mathrm{keV}$, $E_\mcp=1\mathrm{MeV}$ (\ref{fig:sigma1}) and $m_\mcp=1\mathrm{MeV}$, $E_\mcp=1.5\mathrm{MeV}$ (\ref{fig:sigma2}). The photo-absorption cross section is obtained from \cite{Henke:1993eda}. When the millicharged particle is ultra-relativistic, PAI agrees with the maximum between FEA and EPA as expected, when $E_\mcp$ and $m_\mcp$ are relatively close, EPA ceases to be a good approximation, while PAI continues to capture the enhancement at low energy transfer and the free electron behavior at high energy transfer.}
    \label{fig:detection}
\end{figure}

The MCPs entering the detector can interact electromagnetically with the detection material. If the energy exchange is comparable to the binding energy of atoms, an electron could be released from the atom, producing a detectable signal. The use of this atomic ionization process in millicharged particle searches as well as methods to calculate the cross section are discussed in \cite{TEXONO:2018nir,CONNIE:2024off,LZ:2024iwc}.
In the asymptotic regime, when the binding energy and the momentum of the initial electron can be neglected, the cross section for a MCP to transfer at least an energy of $T_0$ to the electron is given by the free electron approximation (FEA), $\sigma(T>T_0) \simeq \varepsilon^2 \times 2.6\times 10^{-25}\,{\rm cm^2}\times ({\rm MeV}/T_0)$ \cite{Magill:2018tbb}.
For the range of $m_\mcp$ and $E_\mcp$ of our interest, the Photo Absorption Ionization model (PAI) is more realistic as it accounts for the asymptotically FEA behavior when the energy transfer is significantly larger than the electron binding energy, as well as for the deviation from FEA when these two energy scales become close. Thus, we adopt PAI model where the differential cross section per atom is given by
\begin{equation}
\begin{aligned}
    \frac{d\sigma}{dT}=&\frac{\varepsilon^2\alpha}{\beta^2\pi}\frac{\sigma_\gamma(T)}{T}\ln\left[\left(1-\beta^2\epsilon_1\right)^2+\beta^4\epsilon_2^2\right]^{-1/2}+\frac{Z\varepsilon^2\alpha}{\beta^2\pi N_e}\left(\beta^2-\frac{\epsilon_1}{\abs{\epsilon}^2}\right)\Theta\\
    &+\frac{\varepsilon^2\alpha}{\beta^2\pi}\frac{\sigma_\gamma(T)}{T}\ln\left(\frac{2m_e\beta^2}{T}\right)+\frac{\varepsilon^2\alpha}{\beta^2\pi}\frac{1}{T^2}\int_0^T\sigma_\gamma(T')dT',
\end{aligned}
\end{equation}
In these expressions,  $T$ is the energy transfer, $\sigma_\gamma$ is the photo-absorption cross section, $\beta$ is the speed of $\mcp$, $\epsilon_1$ and $\epsilon_2$ are the real and imaginary part of the dielectric constant (not to be confused with the value of the millicharge, $\varepsilon$), $N_e$ is the electron number density, and $\Theta=\mathrm{arg}\left(1-\epsilon_1\beta^2+i\epsilon_2\beta^2\right)$. Details of the model can be found in \cite{Allison:1980vw}. A comparison between the PAI model and other models is shown in Fig.\,\ref{fig:detection}.

\begin{figure}[t]
    \centering
    \begin{subfigure}[b]{0.49\textwidth}
    \includegraphics[width=\textwidth]{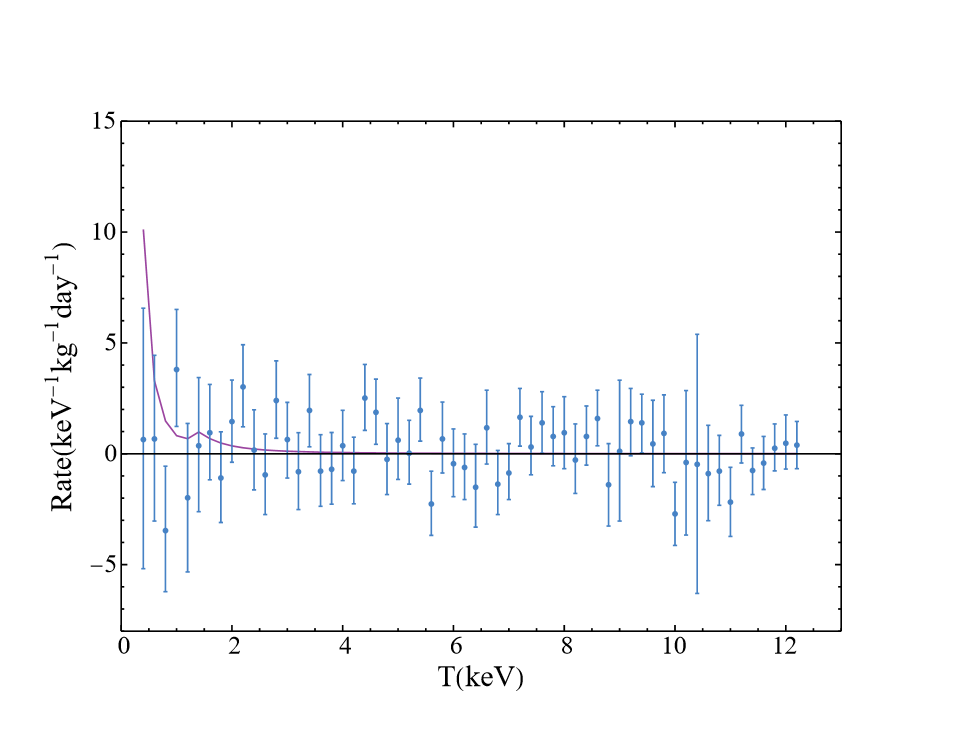}
    \caption{}
    \label{fig:fit}
    \end{subfigure}
    \begin{subfigure}[b]{0.49\textwidth}
    \includegraphics[width=\textwidth]{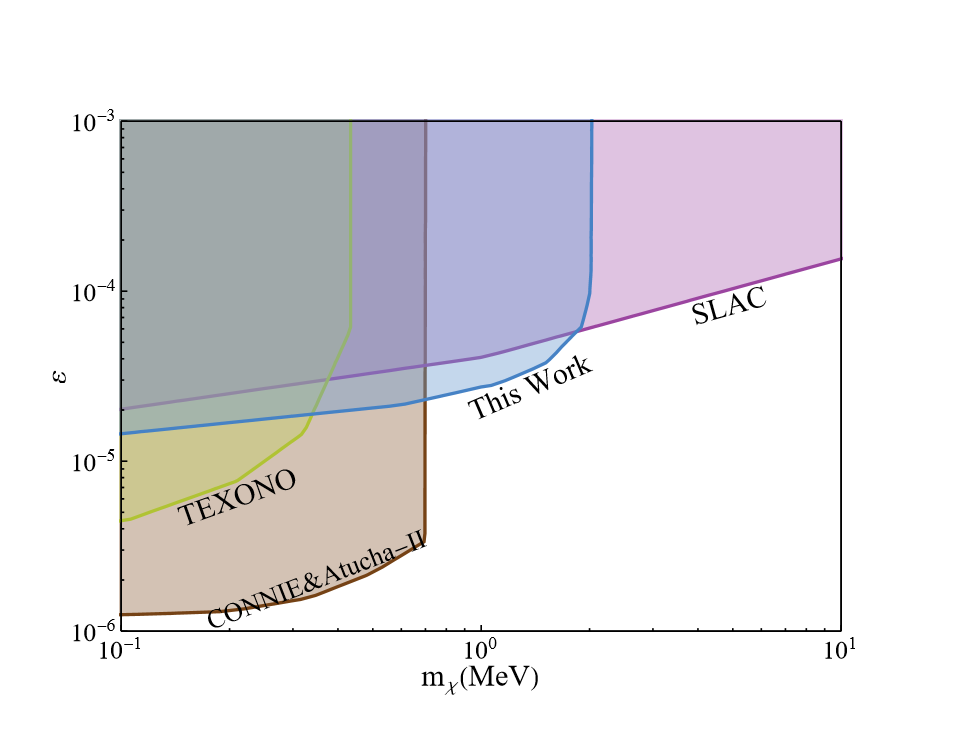}
    \caption{}
    \label{fig:constraint}
    \end{subfigure}
    \caption{(\ref{fig:fit}): An example of the comparison between the analysis from TEXONO (blue dots) and this work (purple line) with $m_\mcp=1\mathrm{MeV}$, $\millicharge=2.7\times10^{-5}$. The constraint on the millicharge is obtained by requiring that the differential counting rate does not exceed the $2\sigma$ region of the TEXONO analysis. Near $T=1.4~ \mathrm{keV}$, the energy transfer becomes enough to ionize electrons in the L shell of Ge (as documented in \cite{Henke:1993eda}) and thus produces a small peak in the rate.  (\ref{fig:constraint}): The constraint we get on MCPs based on comparing the MCP flux generated by nuclear de-excitation process with the TEXONO analysis \cite{TEXONO:2018nir}, in comparison to the constraints from TEXONO \cite{TEXONO:2018nir}, CONNIE and Atucha-II \cite{CONNIE:2024off}, and SLAC \cite{Prinz:1998ua}. Our result, based on theoretical re-evaluation of the MCP flux in conjunction with past experimental searches, provides the strongest constraint in the mass range $0.7-2\mathrm{MeV}$.}
    \label{fig:results}
\end{figure}

The differential count rate is then determined by the MCP flux and the atomic ionization cross section:
\begin{equation}
    \frac{dR}{dT}=\rho_A\int_{E_{\mcp\mathrm{, min}}}^{E_{\mcp\mathrm{, max}}} \frac{d\sigma}{dT}\frac{d\phi_\mcp}{dE_\mcp}dE_\mcp,
    \label{eq:countrate}
\end{equation}
where $\rho_A$ is the number of atoms per unit mass. 
To set the constraint on the millicharge we compare the differential count rate in this work with the experiment and analysis performed by the TEXONO collaboration \cite{TEXONO:2018nir}, which utilizes a PCGe detector with an energy threshold of $300~\mathrm{eV}$ and a mass of $500~\mathrm{g}$ located at $28~\mathrm{m}$ away from a $2.9~\mathrm{GW}$ thermal power nuclear reactor.
The limiting value for $\millicharge$ is determined when the counting rate (\ref{eq:countrate}) is matched to the experimental rate in \cite{TEXONO:2018nir}. An example of this type of comparison is shown in Fig.\ref{fig:fit}. Note that the millicharge is proportional to the fourth root of the count rate, therefore a small uncertainty in the comparison will only lead to a tiny change in the constraint. The constraint we get and comparison with other work at the relevant mass range is shown in Fig.\ref{fig:constraint}.
We observe that the inclusion of nuclear transition enhances the sensitivity to $\varepsilon$, and indeed in the range of $0.7-2$\,MeV produces the strongest constraint to date. 

We discuss in passing a number of other near-reactor experiments with different thresholds and counting rates. The CONNIE and Atucha-II experiments \cite{CONNIE:2024off} utilize Skipper-CCDs, allowing the detection of particles interacting with bulk plasmons at the eV energy scale. For MCP pair-production from photons scattering with electrons, their result shows a better sensitivity compared with TEXONO. On the other hand, for the pair-production from nuclear de-excitation, our calculation shows TEXONO has a slightly better sensitivity, but the details depend on a precise knowledge of the interaction cross-section at low energy transfer. The MUNU experiment \cite{MUNU:2005xnz} tracks electron recoils above 0.7\,MeV with a differential counting rate of $\sim 10^{-1} \mathrm{MeV}^{-1}\mathrm{kg}^{-1}\mathrm{day}^{-1}$ from the hypothetical neutrino magnetic moment, which could as well be from MCPs. Our estimate shows the sensitivity to $\varepsilon^4$ from the MUNU experiment on millicharged particles is a factor of a few hundred weaker than the TEXONO experiment, mostly because of $1/T$ scaling of the cross section. The TEXONO $\bar{\nu}_e$-electron scattering experiment \cite{TEXONO:2009knm} could also in principle receive signals from MCPs, although at a threshold of $T>3\,\mathrm{MeV}$ both the allowed MCP mass range and the available phase space become limited. Finally, this year, the CONUS+ experiment has achieved very low counting rates and very low thresholds, finally detecting the neutrino-nucleus coherent scattering for reactor neutrinos \cite{Ackermann:2025obx}. These results could potentially be also recast as constraints on MCPs.

\section{MCPs from naturally occurring decays and reactions}

Away from the nuclear reactors, excited nuclei are naturally produced from the decay chain of natural radioisotopes. This will result in a geo-background of MCPs detectable by large deep-underground detectors including Borexino \cite{Borexino:2017rsf}, XENON \cite{XENON:2022ltv}, LZ \cite{LZ:2023poo}, and PANDAX \cite{PandaX:2024cic}. Also, $pp$ and CNO reaction chains in the Sun produce photons in the 0.5\,MeV-7.5\,MeV range which should also contribute to the natural MCP abundance.  

The main $\gamma$ emission channels from natural terrestrial radioactivity with energy above $1\,\mathrm{MeV}$ include a number of emission lines in the de-excitation of $^{214}\mathrm{Bo}$ in the $^{238}\mathrm{U}$ chain, $^{208}\mathrm{Pb}$ in the $^{232}\mathrm{Th}$ chain, and $^{40}\mathrm{Ar}$ in the $^{40}\mathrm{K}$ decay. Especially promising are the $2.615\,\mathrm{MeV}$ E3 emission line from the de-excitation of $^{208}\mathrm{Pb}$ occurring at the highest energy, and the $1.461\,\mathrm{MeV}$ E2 emission line in $^{40}\mathrm{K}$ decay accompanied by the highest component of the geo-neutrino flux \cite{huang2013reference}\footnote{More nuclear data resources can be found at website http://www.lnhb.fr/home/nuclear-data/nuclear-data-table/}.

To estimate the MCP flux, we focus on the XENONnT experiment \cite{XENON:2022ltv} and normalize the MCP flux on the geo-neutrino flux from the same decay chain measured by the Borexino experiment \cite{Borexino:2015ucj} at the same site.  Different from neutrinos which travel through the Earth almost freely, MCPs lose energy as they travel through the Earth as a result of the electromagnetic interaction with the surrounding material, based on the review in \cite{ParticleDataGroup:2024cfk}, the energy loss rate for typical values of $\beta\gamma$ under consideration for MCPs travelling through the Crust is estimated to be
\begin{equation}
    \frac{dE}{dx}=-\frac{(0.003-0.03)\,\mathrm{MeV}}{100\,\mathrm{km}}\left(\frac{\millicharge}{10^{-5}}\right)^2,
\end{equation}
therefore, for the values of $\millicharge$ of our interest, it is appropriate to focus on the MCPs produced from the Local Crust (LOC, within $\sim$ 300\,km from the detector, see \cite{huang2013reference} for details), where the energy loss of MCPs can be neglected. Based on the geo-neutrino measurement available at Borexino, we focus specifically on the MCP pair production accompanying the $2.615\,\mathrm{MeV}$ emission line in the $^{232}\mathrm{Th}$ chain. The geo-MCP flux is then determined by
\begin{equation}
    \frac{d\phi_{\mcp}}{dE_{\mcp}}=\frac{2}{6}\frac{dr_{\mathrm{E3}}}{dE_{\mcp}}\phi_{\nu}(^{232}\mathrm{Th},\mathrm{LOC}),
\end{equation}
where the prefactor $2/6$ comes from one $^{208}\mathrm{Pb}$ de-excitation per six neutrino emission in the $^{232}\mathrm{Th}$ chain and two MCPs per pair production; ${dr_{\mathrm{E3}}}/{dE_{\mcp}}$ is the ratio between MCP pair production events and $\gamma$ emission events for E3 multipole transition, with the explicit form derived in Appendix \ref{multipole pair production}. Finally, $\phi_{\nu}(^{232}\mathrm{Th},\mathrm{LOC})$ denotes the neutrino flux from $^{232}\mathrm{Th}$ decay in the Local Crust. Assuming that the $^{238}\mathrm{U}/^{232}\mathrm{Th}$ abundance ratio is independent of the material considered, $\phi_{\nu}(^{232}\mathrm{Th},\mathrm{LOC})$ corresponds to approximately $\sim10\%$ of the total geo-neutrino flux measured by the Borexino experiment.

Under the above assumption, the constraint we get is $\sim\millicharge<10^{-4}$ at $m_\mcp\sim 1\,\mathrm{MeV}$, however, at this value of $\millicharge$ the energy loss in the Local Crust becomes non-eligible, therefore, we were not able to set a solid independent bound on $\millicharge$ based on the XENONnT experiment. We also considered the Borexino experiment which has a larger detector volume, but due to the higher detection threshold the MCP slow-down is even more crucial, and the conclusion is similar.

It is well appreciated that the total solar neutrino flux dominates over geo-neutrino $\bar\nu_e$ flux by $\sim$ four orders of magnitude. Using well-understood properties of the solar energy generation cycles, we can easily calculate the MCP production rate. Of course, for the $O(10^{-5})$ values of the millicharge, the $\chi$ particles produced in the core of the Sun will be thermalized and trapped in the solar interior for considerable time. The main processes that will create $\chi's$ are listed below: 
\begin{eqnarray}
    e^++e^- &\to& \chi + \bar\chi;~~m_\chi^{max} = 0.511 \,{\rm MeV}, \label{pp}\\
    {\rm D} + p &\to& {\rm ^3He}+ \chi + \bar\chi;~~2m_\chi^{max} = 5.5 \,{\rm MeV}, \label{Dp}\\
    {\rm ^3He}+{\rm ^3He}&\to& {\rm ^6Be}+\chi + \bar\chi;~~2m_\chi^{max} = 11.5 \label{Be6} \,{\rm MeV}, \\
    {\rm ^{12}C}+p&\to& {\rm ^{13}N}+\chi + \bar\chi;~~2m_\chi^{max} = 1.9 \,{\rm MeV} \label{C12} \\
    {\rm ^{13}C}+p&\to& {\rm ^{14}N}+\chi + \bar\chi;~~2m_\chi^{max} = 7.5 \,{\rm MeV} \label{CNO}\\
    {\rm ^{14}N}+p&\to& {\rm ^{15}O}+\chi + \bar\chi;~~2m_\chi^{max} = 7.3 \,{\rm MeV} .
\end{eqnarray}
In addition, there are processes with the de-excitations of $^7{\rm Be}$ and $^7{\rm Li}$ that create 0.43 and 0.48\,MeV energy pairs. 
The corresponding photon-emitting cross sections for most of the listed reactions are relatively well-known. Before evaluating the rate of MCP production, we would like to make several remarks:
\begin{itemize}
    \item One positron is emitted for each unit of $pp$ cycle. $e^+$ annihilation is an important source of MCP, because in that case the probability of $\chi$ production scales as $O(\varepsilon^2)$, and not $O(\alpha \varepsilon^2)$.

\item Process (\ref{Be6}), ${\rm ^3He}+{\rm ^3He}$ to ${\rm ^6Be}$, is not very well known as the transition occurs due to E0 multipole and should be suppressed. This reaction is expected to be subdominant as the main ${\rm ^3He}+{\rm ^3He}$ fusion reaction occurs without photon emission.

\item N$(p,\gamma)$O reaction to the ground state of oxygen is suppressed (Br $\sim$ 6\%) and can be neglected. Reaction (\ref{CNO}), on the other hand, occurs mostly due to E1 transition to the ground state (see {\em e.g.} \cite{Ajzenberg-Selove:1991rsl} and TUNL Nuclear Data Project compilations\footnote{~https://nucldata.tunl.duke.edu/}). Reaction (\ref{C12}) can be neglected on account of small $Q$-value and the overall subdominance of the CNO chain.

\item We shall assume that D$+p$ fusion reaction at solar energies is an equal mix of E1 and M1 transitions with more advanced treatment possible if needed \cite{Marcucci:2015yla}. This assumption is not crucial if $2m_\chi $ is away from $ 5.5$\,MeV threshold. 

\item MCPs can also be emitted at order $O(\alpha^2\varepsilon^2)$ from weak decays of {\em e.g.} ${\rm ^8B}$. These are even smaller contributions than the rest, and we neglect them. 
    
\end{itemize}

\begin{figure}[t]
    \centering
\includegraphics[width=0.65\textwidth]{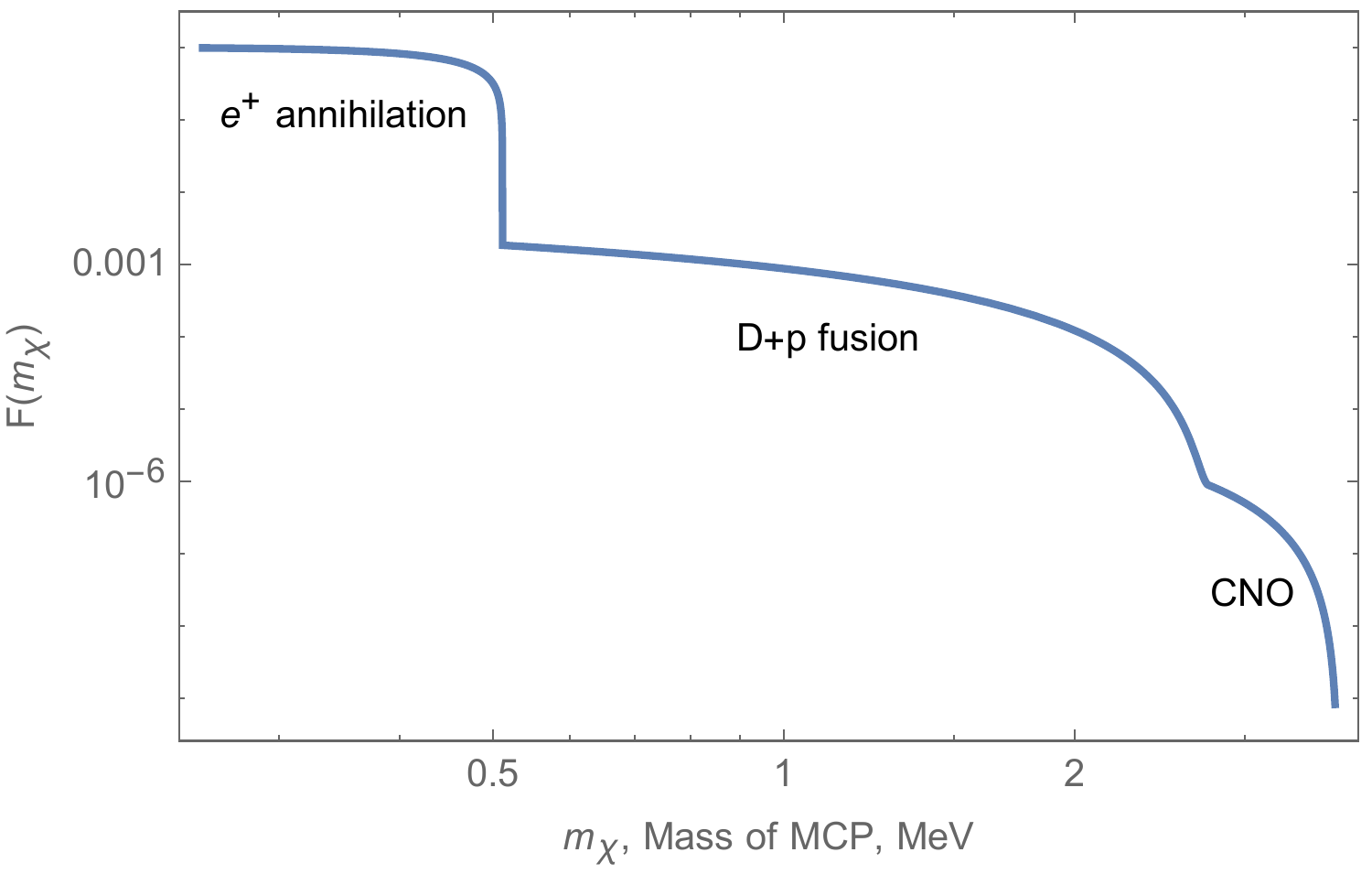}
    \caption{Relative production of solar MCPs due to different reactions. For $m_\chi < m_e$ the flux is dominated by the positron annihilation.}
    \label{fig:solarMCP}
\end{figure}

  In light of the above, we use reactions (\ref{pp}), (\ref{Dp}) and (\ref{CNO}) together with our results for the branching ratios into $\chi\bar\chi$ to calculate the overall production rate as function of $m_\chi$. A convenient way of expressing the answer is as follows: 
\begin{eqnarray}
\label{wouldbeflux}
    \frac{1}{4\pi({\rm A.U.})^2} \frac{dN_{\chi+\bar\chi}}{dt} \simeq 12\,{\rm cm^{-2}s^{-1}}\times \left(\frac{\varepsilon} {10^{-5}}\right)^2\times F(m_\chi),
\end{eqnarray}
where the function $F(m_\chi)$, plotted in Figure \ref{fig:solarMCP}, is dictated by the available phase space, as well as the relative strength of $pp$ and CNO chains. Notice that the probability to create an MCP pair from the positron annihilation is given by 
\begin{eqnarray}
    \frac{\langle \sigma v \rangle_{e^+e^-\to \chi\bar\chi}}{\langle \sigma v \rangle_{e^+e^-\to \gamma\gamma}} = \varepsilon^2\times \left(1+\frac{r^2}{2}\right)\sqrt{1-r^2},~~r = m_\chi/m_e. 
\end{eqnarray}
In the limit of $m_\chi\ll m_e$, $F=1$, and this probability is just $\varepsilon^2$.  Therefore the total MCP flux is $2\varepsilon^2$ times the $pp$ neutrino flux, which explains the prefactor in Eq. (\ref{wouldbeflux}). The discussion of production of other exotic particles in solar nuclear reactions can be found in recent work \cite{DEramo:2023buu}.

Quantity (\ref{wouldbeflux}) is not affected by the slow-down of millicharged particles by the solar material, and in the steady state regime will correspond to an outgoing flux at an astronomical unit away from the Sun. The average energy of particles arriving from the Sun is, however, a more difficult question. One expects that for $\varepsilon>10^{-6}$ most of the MCPs will thermalize and leave the Sun with relatively small energies. In that case, their propagation near the Earth's surface may get affected by the magnetic fields. On the other hand, MCP particles with $\varepsilon< 10^{-7}$ are expected to exit the Sun without appreciable energy loss. They can be then very efficiently probed by the planned low electron recoil experiments such as Oscura \cite{Oscura:2023qch}. We note that the recent results from the DAMIC-M experiment \cite{DAMIC-M:2025luv} have a potential sensitivity to $\varepsilon \sim 0.5 \times 10^{-6}$ via the flux of solar MCPs. However, more in-depth investigation of $\chi$ slow-down is needed to fully address this question. Our calculation of the overall flux (\ref{wouldbeflux}) can be further augmented by the secondary Compton-like production of MCPs $e+\gamma\to e+ \chi+\bar\chi$ from all photons produced in nuclear reactions. 

\section{Dark photon from nuclear de-excitation}

In addition to MCPs the nuclear de-excitation could also produce massive dark photons $A'$ which effectively interact with ordinary matter through the coupling $\kappa Q J^\mu A'_\mu$. Following the formalism in \cite{Pitrou:2019pqh} and accounting for the dark photon mass, we obtain the ratio between dark photon and regular photon production rate for E1 and M1 transition as

\begin{align}
    &r_{\mathcal{E}}^{(A')}=\kappa^2 \left(1+\frac{1}{2}\frac{m_{A'}^2}{\omega^2}\right)\sqrt{1-\frac{m_{A'}^2}{\omega^2}},\\
    &r_{\mathcal{B}}^{(A')}=\kappa^2\left(1-\frac{m_{A'}^2}{\omega^2}\right)^{\frac{3}{2}}.
\end{align}

\begin{figure}[t]
    \centering
    \includegraphics[width=0.5\linewidth]{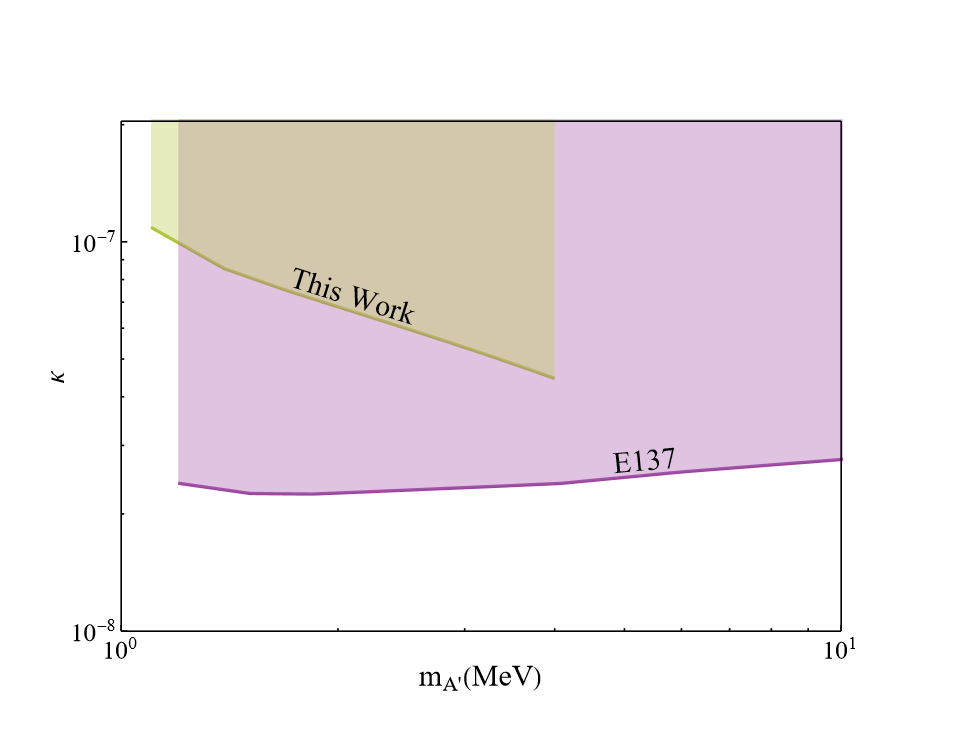}
    \caption{Excluded region for dark photon from this work and the SLAC E137 experiment \cite{Bjorken:1988as,Andreas:2012mt,Marsicano:2018krp}. The entire plot region is also excluded by the constraint based on supernova \cite{Chang:2016ntp}.}
    \label{fig:darkphoton}
\end{figure}

The produced dark photons subsequently decay to an electron-positron pair, allowing the possibility for detection. The decay rate of dark photons is given by
\begin{equation}
    \Gamma=\frac{\alpha\kappa^2m_{A'}}{3}\left(1+\frac{2m_e^2}{m_{A'}^2}\right)\sqrt{1-\frac{4m_e^2}{m_{A'}^2}}.
\end{equation}
For MeV scale dark photon with $\kappa\sim 10^{-7}$, the typical distance $L=\gamma\beta/\Gamma$ dark photons travel during their lifetime is a few kilometers, and at that distance the geo-dark photons are unlikely to play any role. Therefore we focus solely on detectors that are placed at a distance $D\ll L$ from the nuclear reactors. Since the loss of dark photons from the source to the detector can be ignored, the dark photon decay event rate inside the detector is given by
\begin{equation}
    R=\phi_{A'}V/L,
\end{equation}
where $\phi_{A'}=N_\gamma^{(N,\omega)}r_{\mathcal{E}(\mathcal{B})}^{(A')}/(4\pi D^2)$ is the dark photon flux, and $V$ is the detector volume. Given the proportionality of the event rate to the detector volume, dark photons are ideally constrained by large size detectors near nuclear reactors. We considered the measurement by the TEXONO Collaboration with CsI(Tl) scintillating crystals \cite{TEXONO:2009knm}, that measured $\bar\nu_e$ scattering on electrons.  A dark photon decay inside the detector would be classified as a three-hit cosmic-ray unrelated event, and the counting rate that was observed for this class of events was at the level of $O(10^{-2}\,{\rm events/kg/day/MeV})$.  Somewhat less competitive limits can be derived from Ref. \cite{Hagner:1995bn}, that searched for the appearance of $e^+e^-$ pairs from a hypothetical massive neutrino decay. Among the emission lines under consideration, the $4.060\mathrm{MeV}$ emission line from ${\rm ^{238}U}(n,\gamma){\rm ^{239}U}$ provide the strongest sensitivity. The constraint we get by repurposing results of Ref. \cite{TEXONO:2009knm}, together with comparison with other constraints at the same mass range, are shown in Fig.\,\ref{fig:darkphoton}. While our constraint does not exceed existing limits, the use of reactors as the source of dark photon provides an independent confirmation for the exclusion of dark photons at this mass and coupling range.

One can also set constraints on visibly decaying dark photons using large neutrino detectors deep underground. Specifically, KamLAND detector measured the solar $^8{\rm B}$ neutrinos, achieving effectively $\sim5\,\rm MeV$ detection threshold \cite{KamLAND:2011fld}. A dark photon produced in nuclear reactors with $E_{A'}>5\,\rm MeV$ will appear in that sample, and therefore can be constrained {\em if} energetic enough nuclear transitions can be found. The lines considered in this paper up to this point are not suitable as they would be below the threshold. One candidate for more energetic transitions is the neutron capture on iron, $^{56}{\rm Fe}(n,\gamma)^{57}{\rm Fe}$, that does result in significant emission of $7.6\,\rm MeV$ gammas (corresponding to two fairly strong M1 transitions \cite{Firestone:2017fet}, discussed recently in connection with axion searches \cite{Waites:2022tov}). It is somewhat difficult to estimate exact fraction of neutrons captured on reactor walls without detailed simulations, and as a rough estimate we adopt $1\%$. With these numbers in mind, and taking and average baseline of 180\,km, and the total thermal power of all reactors to be $\sim 100\,\rm GW$, we can calculate the expected event rate from $A'$ decay and compare it with recorded event rate in KamLAND in the electron energy bin $7.5-8$\,MeV. In this energy range, most of the observed signal is due to the solar neutrinos. For simplicity, we quote our result for a specific value of the dark photon mass of 2\,MeV. 
\begin{eqnarray}
    \left. \frac{dN_{A'\,\rm decay}/dt}{dN_{\rm observed}/dt}\right|_{m_{A'}= 2\,\rm MeV} \sim 4\times 10^{-4} \times \left(\frac{\kappa}{3\times 10^{-8}}\right)^4 .
\end{eqnarray}
Notice that the normalization of $\kappa$ in this expression is chosen to coincide with the decay distance similar to KamLAND baseline, as larger $\kappa$ would lead to a dominant fraction of $A'$ decay before the detector. 
We conclude that the sensitivity of a ``far detector" is less stringent than the results of the experiments near the reactor cores. It is possible that in the near future the JUNO experiment \cite{JUNO:2015zny} could surpass KamLAND and set tighter bounds on visibly decaying $A'$. 

We would also like to discuss a special case of $m_{A'} < 2 m_e$. In this case, the visible decay mode is a loop-induced $A'\to 3\gamma$ \cite{Pospelov:2008jk,McDermott:2017qcg}, and it  is very suppressed. The main mechanism for detecting $A'$ is in fact a Compton-like scattering, $e+A'\to e+\gamma$. Using this process, Ref. \cite{Pospelov:2017kep} derived novel limits on dark photon in this mass range, obtaining $\kappa<10^{-5}$. Would nuclear reactors be able to improve on this bound? The inverse Compton process will create $e\gamma$ in the final state, and it is not obvious what the limits are on this sub-category of events from the existing searches in detectors near the reactor cores. It will definitely have background from similar channels in $^{208}\rm Tl$ decays. To avoid this complication, we use KamLAND signal again, and estimate that 
\begin{eqnarray}
    \left. \frac{dN_{e+A'\to e+\gamma}}{dN_{\rm observed}/dt}\right|_{m_{A'}<2m_e} \sim 0.1 \times \left(\frac{\kappa}{3\times 10^{-5}}\right)^4 .
\end{eqnarray}
As before, neutron capture on iron nuclei is used as a source of $A'$. We conclude that currently the reactors provide subdominant sensitivity to the LSND limits for dark photons below $2m_e$ mass. JUNO, however, would be able to improve this limit significantly, mostly due to a superb energy resolution and the possibility to look for sharp lines in the spectrum, and from the $\sim 3.5$ times smaller baseline. 
 
\section{Discussion and Conclusions}

Experiments at nuclear reactor experiments have been instrumental in determining the properties of neutrinos and constraining new physics such as axions and axion-like particles. Clearly, other MeV-scale particles from dark sectors can be constrained as well, with clear prospects of improving sensitivity to several classes of models (for an axion-like model see {\em e.g.} recent works \cite{Brdar:2020dpr,NEON:2024kwv}, and for very light dark photons \cite{Danilov:2018bks}). In this paper, we have revisited the constraints on millicharged particles imposed by detectors placed very near reactor cores. We have also  calculated the production of MCPs from natural radioactivity and nuclear reactions powering the Sun. The main difference with the previous treatment is that we point out that every photon emission in nuclear $\gamma$ decays can be accompanied by the $\chi\bar\chi$ production provided that $2m_\chi<\omega$. We have derived the corresponding energy spectrum of MCP particles, using previous simulations of the $\gamma$ emission, mostly from $(n,\gamma)$ reactions and decays of radioactive fission products. The calculated flux is typically much harder than the Compton-type production of MCP in the $e\gamma$ scattering. Consequently, higher $m_\chi$ masses can be probed, and we derive novel limits in the $0.7-2$ MeV range, surpassing the sensitivity of the SLAC MCP search. These limits can be further improved, as the MCP scattering cross section scales inversely proportional to the electron recoil energy, and new low-threshold detectors will help to increase sensitivity. 

While nuclear reactors provide the largest fluxes of MCPs for the closely located detectors, we note that there is also an ambient flux of MCP created by natural radioactivity (mostly caused by the radioactive isotopes of K, U and Th). While fluxes of MCP created this way are much lower than those in proximity of reactor cores, the detection methods that can be employed could involve neutrino detectors, and large self-shielding dark matter detectors (such as those based on xenon dual-TPC technology). These are large deep-underground detectors operating in a maximally background-free environment, capable of achieving much lower counting rates than detectors placed in reactors' proximity. While a solid constraint on $\millicharge$ based on these detectors relies on a detailed knowledge of the composition of the Local Crust, our analysis shows that the constraints derived in this way are likely to fall within the ballpark of $\millicharge\sim 10^{-4}$. 

Perhaps more promising avenue for the future studies is the solar MCP flux. If $m_\chi$ is below the $m_e$ the MCPs will be very efficiently produced by the positron annihilation. For heavier MCPs, the reaction ${\rm D(}p,\gamma){\rm ^3He}$ will provide the dominant contribution up to $m_\chi \sim 2.5$\,MeV. Future low recoil experiments will be sensitive to the flux even for a very small value of millicharge, $\varepsilon < 10^{-6}$.

The improved energy reach achieved by considering the $(n,\gamma)$ reactions also allows the on-shell production of dark photons heavy enough to decay to $e^+e^-$ pairs. We derived the rate for near-reactor detectors to detect such decay signals and converted the observed signal rate to the constraints on dark photon couplings in the $1.1-4$ MeV range. While the constraints derived in this way do not supplant the existing limits, we note that this is the first dark photon constraint obtained from near-reactor experiments for the MeV range, and the limits can be further improved with a detailed analysis of the background and the use of larger volume detectors.

Before we close, we would like to point to several items of discussion:
\begin{itemize}

    \item We improved the reactor bounds on MCPs because $(n,\gamma)$ reactions give access to higher frequencies and higher $m_\chi$ masses compared to bremsstrahlung pair-production of $\chi\bar\chi$. It would be also important to investigate whether in the $m_\chi< m_e$ range the bounds could be improved even further. A considerable fraction of these hard $\gamma$'s would undergo an $e^+e^-$ pair production on heavy nuclei such as uranium. Subsequent annihilation of $e^+$ with a branching to MCPs would produce $\chi$'s with the $O(\varepsilon^2)$ probability, which is much larger than $O(\alpha\varepsilon^2)$ in case of the bremsstrahlung production. In order to be able to do this re-analysis, one would have to simulate the relative weight of the pair production and Compton degradation of hard $\gamma$, or in other words, determine the yield of $e^+$. This will be best done by experimental collaborations themselves, accounting for photon conversions inside (possibly dominated by pair production) and outside (presumably dominated by Compton process) nuclear fuel rods. 

    \item In light of the recent success in developing extremely low threshold DM detectors \cite{DAMIC-M:2025luv}, one would need a better quality estimate of the solar MCP flux. More careful calculations will take into account the spatial distribution of $\chi\bar\chi$ pair-production inside the solar core, and most importantly the subsequent slow-down, and, for large enough $\varepsilon$, eventual thermalization of these particles. Estimates show that $\varepsilon\sim 10^{-6}$ appears to be a critical value of the millicharge, but a better calculation is feasible. 

    \item The range of masses and couplings explored in this paper is relevant for the possible solutions of EDGES anomaly. One should not, however, loose sight of the fact that such light MCPs are troublesome from the point of the very early cosmology, as they would certainly modify the predictions of the Big Bang Nucleosynthesis (BBN). ``Fixing" deuterium and helium abundances will most likely require additional ingredients/actors from the DS at the time of BBN. 

    \item While $(n,\gamma)$ reactions explored in this paper are necessarily limited in their kinematic reach to $2m_\chi$ and/or $m_{A'}$ under 5 MeV, the question remains of how efficiently the process of direct nuclear fission, {\em i.e.} nuclear scission itself, may produce exotic particles. While it will be suppressed by some powers of $v/c$, where $v$ is the velocity of separating fragments, the potential kinematic reach will be higher than for the $(n,\gamma)$ sources, as up to 180 MeV is available per every single occurrence of fission. 

    \item It is worth noting that deep underground far detector can also be used for deriving constraints on visibly decaying $A'$. It remains to be seen if the ongoing JUNO program will improve on some of the discussed limits, once the solar $^8\rm B$ neutrino signal (suitable for deriving such limits) is measured.

\end{itemize}

\section{Acknowledgments}
The authors would like to thank Drs. A Berlin and Z. Liu for useful discussions. This work is supported in part by the DOE grant DE-SC0011842. MP would also like to acknowledge the hospitality of Perimeter Institute for Theoretical Physics in Waterloo, Ontario, Canada. 

\appendix

\section{Pair production from E2 and E3 transitions}
\label{multipole pair production}
In this appendix we derive the ratio between MCP pair production and $\gamma$ emission rates in nuclear de-excitation for E2 and E3 transitions under the recoil-less approximation. Up to some normalization for the multipole moments we expand the electric part of the nuclear transition current as
\begin{align}
    &J_i=\omega Q^{(1)}_i+\omega Q^{(2)}_{ij}q^j+\omega Q^{(3)}_{ijk}q^jq^k,\\
    &J_0=q_iJ_i/\omega=Q^{(1)}_iq^i+Q^{(2)}_{ij}q^iq^j+Q^{(3)}_{ijk}q^iq^jq^k,
\end{align}
where $Q^{(1)}_i$, $Q^{(2)}_{ij}$, $Q^{(3)}_{ijk}$ are the electric dipole, quadrupole, and octapole moments, and $k=(\omega, \boldsymbol{q})$ is the photon's (or virtual photon's) four-momentum. Evaluating the squared amplitude for pair production and $\gamma$ emission involves products of the form $J_\mu J_\nu^*$, when averaged over all directions and spins, such products can only depend on the magnitude of the multipoles rather than their components, therefore the products of the multipoles can be replaced by
\begin{align}
    Q^{(1)i}Q^{(1)*}_j&\to \frac{1}{3} \left\vert Q^{(1)}\right\vert^2\delta^i_j,\\
    Q^{(2)ij}Q^{(2)*}_{kl}&\to\frac{1}{5} \left\vert Q^{(2)}\right\vert^2\left(\delta^{(i}_{(k}\delta^{j)}_{l)}-\frac{1}{3}\delta^{(ij)}\delta_{(kl)}\right),\\
    Q^{(3)ijk}Q^{(3)*}_{lmn}&\to\frac{1}{7} \left\vert Q^{(3)}\right\vert^2\left(\delta^{(i}_{(l}\delta^{j}_{m}\delta^{k)}_{n)}-\frac{3}{5}\delta^{(ij}\delta^{k)}_{(l}\delta_{mn)}\right),\\
    Q^{(m)i\cdots j}Q^{(n)*}_{k\cdots l}&\to 0, \quad \mathrm{for }\quad  m\neq n,
\end{align}
where the coefficients are determined by the requirement $Q^{(m)i\cdots j}Q^{(m)*}_{i\cdots j}=\left\vert Q^{(m)}\right\vert^2$ and the traceless condition of the multipole moments. These results then lead to
\begin{align}
    J_0J_0^*&\to \frac{1}{3}\left\vert Q^{(1)}\right\vert^2q^2+\frac{2}{15}\left\vert Q^{(2)}\right\vert^2 q^4+\frac{2}{35}\left\vert Q^{(3)}\right\vert^2q^6,\\
    J_0J_i^*&\to \frac{1}{3}\left\vert Q^{(1)}\right\vert^2\omega q_i+\frac{2}{15}\left\vert Q^{(2)}\right\vert^2 \omega q^2q_i+\frac{2}{35}\left\vert Q^{(3)}\right\vert^2\omega q^4q_i,\\
    J_iJ_j^*&\to\frac{1}{3} \left\vert Q^{(1)}\right\vert^2\omega^2\delta_{ij}+\frac{2}{15}\left\vert Q^{(2)}\right\vert^2 \omega^2\left(\frac{3}{4}\delta_{ij}q^2+\frac{1}{4}q_iq_j\right)+\frac{2}{35}\left\vert Q^{(3)}\right\vert^2 \omega^2q^2\left(\frac{2}{3}\delta_{ij}q^2+\frac{1}{3}q_iq_j\right).
\end{align}
For $\gamma$ emission the squared amplitude is given by
\begin{equation}
    \abs{\mathcal{M}}^2_{(\gamma)}=-g^{\mu\nu}J_{\mu}J_{\nu}^*\to \frac{2}{3}\left\vert Q^{(1)}\right\vert^2 q^2+\frac{1}{5}\left\vert Q^{(2)}\right\vert^2 q^4+\frac{8}{105}\left\vert Q^{(3)}\right\vert^2 q^6,
\end{equation}
for pair production the squared amplitude is given by
\begin{equation}
    \abs{\mathcal{M}}^2_{(\mcp\mcpb)}=T^{\mu\nu}J_{\mu}J_{\nu}^*,\qquad T^{\mu\nu}=\frac{\millicharge^2e^2}{k^4}\left(-g^{\mu\nu}2k^2+4p_\mcp^\mu p_\mcpb^\nu+4p_\mcpb^\mu p_\mcp^\nu\right),
\end{equation}
after some algebra we arrive at
\begin{align}
    \abs{\mathcal{M}}^2_{(\mcp\mcpb,\mathrm{E1})}&=\frac{2e^2\millicharge^2}{3k^4}\left\vert Q^{(1)}\right\vert^2\left[k^4+k^2\left(-4E_\mcp E_\mcpb +2\omega^2\right)+4m_\mcp^2\omega^2\right],\\
    \abs{\mathcal{M}}^2_{(\mcp\mcpb,\mathrm{E2})}&=\frac{e^2\millicharge^2}{15k^4}\left\vert Q^{(2)}\right\vert^2\left[-4k^6+k^4\left(16E_\mcp E_\mcpb -3\omega^2\right)\right.\nonumber\\
    &\left.\hspace{3cm}+6k^2\omega^2\left(-2E_\mcp E_\mcpb -2m_\mcp^2+\omega^2\right)+12m_\mcp^2\omega^4\right],\\
    \abs{\mathcal{M}}^2_{(\mcp\mcpb,\mathrm{E3})}&=\frac{4e^2\millicharge^2}{105k^4}\left\vert Q^{(3)}\right\vert^2\left[3k^8-k^6\left(12E_\mcp E_\mcpb +\omega^2\right)+2k^4\omega^2\left(10E_\mcp E_\mcpb +4m_\mcp^2-3\omega^2\right)\right.\nonumber\\
    &\left.\hspace{3cm}+4k^2\omega^4\left(-2E_\mcp E_\mcpb -4m_\mcp^2+\omega^2\right)+8m_\mcp^2\omega^6\right].
\end{align}
To obtain the ratio of cross sections, we integrate over the phase space for each process. This gives the differential ratio in the form of Eq.~\eqref{differentialratio}. After changing variables using $k^2=M^2+M'^2-2ME_{N'}$, the problem reduces to calculating the elementary integral
\begin{equation}
    I_n=\int_{k^2_{\mathrm{min}}}^{k^2_{\mathrm{max}}}k^{2n}dk^2,
\end{equation}
the kinematical limits correspond to $\mcp$ and $\mcpb$ moving parallel or antiparallel to each other, with
\begin{equation}
    k^2_{\mathrm{min}}=2E_\mcp E_\mcpb +2m_\mcp^2-2\abb{p_\mcp}\abb{p_\mcpb}, \quad k^2_{\mathrm{max}}=2E_\mcp E_\mcpb +2m_\mcp^2+2\abb{p_\mcp}\abb{p_\mcpb}.
\end{equation}
Finally, in terms of $I_n$, the differential ratio is given by
\begin{align}
    \frac{dr_{\mathrm{E1}}}{dE_\chi}&=\frac{\alpha\millicharge^2}{4\pi\omega^3}\left[I_0+\left(-4E_\mcp E_\mcpb +2\omega^2\right)I_{-1}+4m_\mcp^2\omega^2I_{-2}\right],\\
    \frac{dr_{\mathrm{E2}}}{dE_\chi}&=\frac{\alpha\millicharge^2}{12\pi\omega^5}\left[-4I_{1}+\left(16E_\mcp E_\mcpb -3\omega^2\right)I_{0}+6\omega^2\left(-2E_\mcp E_\mcpb -2m_\mcp^2+\omega^2\right)I_{-1}+12m_\mcp^2\omega^4I_{-2}\right],\\
    \frac{dr_{\mathrm{E3}}}{dE_\chi}&=\frac{\alpha\millicharge^2}{8\pi\omega^7}\left[3I_{2}-\left(12E_\mcp E_\mcpb +\omega^2\right)I_{1}+2\omega^2\left(10E_\mcp E_\mcpb +4m_\mcp^2-3\omega^2\right)I_{0}\right.\nonumber\\
    &\left.\hspace{3cm}+4\omega^4\left(-2E_\mcp E_\mcpb -4m_\mcp^2+\omega^2\right)I_{-1}+8m_\mcp^2\omega^6I_{-2}\right].
\end{align}

\small
\bibliographystyle{utphys}
\bibliography{ref}

\providecommand{\href}[2]{#2}\begingroup\raggedright\begin{thebibliography}{10}

\bibitem{Lanfranchi:2020crw}
G.~Lanfranchi, M.~Pospelov, and P.~Schuster, ``{The Search for Feebly
  Interacting Particles},''
  \href{http://dx.doi.org/10.1146/annurev-nucl-102419-055056}{{\em Ann. Rev.
  Nucl. Part. Sci.} {\bfseries 71} (2021) 279--313},
  \href{http://arxiv.org/abs/2011.02157}{{\ttfamily arXiv:2011.02157
  [hep-ph]}}.

\bibitem{Beacham:2019nyx}
J.~Beacham {\em et~al.}, ``{Physics Beyond Colliders at CERN: Beyond the
  Standard Model Working Group Report},''
  \href{http://dx.doi.org/10.1088/1361-6471/ab4cd2}{{\em J. Phys. G} {\bfseries
  47} no.~1, (2020) 010501}, \href{http://arxiv.org/abs/1901.09966}{{\ttfamily
  arXiv:1901.09966 [hep-ex]}}.

\bibitem{Holdom:1985ag}
B.~Holdom, ``{Two U(1)'s and Epsilon Charge Shifts},''
  \href{http://dx.doi.org/10.1016/0370-2693(86)91377-8}{{\em Phys. Lett. B}
  {\bfseries 166} (1986) 196--198}.

\bibitem{Davidson:2000hf}
S.~Davidson, S.~Hannestad, and G.~Raffelt, ``{Updated bounds on millicharged
  particles},'' \href{http://dx.doi.org/10.1088/1126-6708/2000/05/003}{{\em
  JHEP} {\bfseries 05} (2000) 003},
  \href{http://arxiv.org/abs/hep-ph/0001179}{{\ttfamily arXiv:hep-ph/0001179}}.

\bibitem{Vogel:2013raa}
H.~Vogel and J.~Redondo, ``{Dark Radiation constraints on minicharged particles
  in models with a hidden photon},''
  \href{http://dx.doi.org/10.1088/1475-7516/2014/02/029}{{\em JCAP} {\bfseries
  02} (2014) 029}, \href{http://arxiv.org/abs/1311.2600}{{\ttfamily
  arXiv:1311.2600 [hep-ph]}}.

\bibitem{Dolgov:2013una}
A.~D. Dolgov, S.~L. Dubovsky, G.~I. Rubtsov, and I.~I. Tkachev, ``{Constraints
  on millicharged particles from Planck data},''
  \href{http://dx.doi.org/10.1103/PhysRevD.88.117701}{{\em Phys. Rev. D}
  {\bfseries 88} no.~11, (2013) 117701},
  \href{http://arxiv.org/abs/1310.2376}{{\ttfamily arXiv:1310.2376 [hep-ph]}}.

\bibitem{Chang:2018rso}
J.~H. Chang, R.~Essig, and S.~D. McDermott, ``{Supernova 1987A Constraints on
  Sub-GeV Dark Sectors, Millicharged Particles, the QCD Axion, and an
  Axion-like Particle},'' \href{http://dx.doi.org/10.1007/JHEP09(2018)051}{{\em
  JHEP} {\bfseries 09} (2018) 051},
  \href{http://arxiv.org/abs/1803.00993}{{\ttfamily arXiv:1803.00993
  [hep-ph]}}.

\bibitem{Dunsky:2018mqs}
D.~Dunsky, L.~J. Hall, and K.~Harigaya, ``{CHAMP Cosmic Rays},''
  \href{http://dx.doi.org/10.1088/1475-7516/2019/07/015}{{\em JCAP} {\bfseries
  07} (2019) 015}, \href{http://arxiv.org/abs/1812.11116}{{\ttfamily
  arXiv:1812.11116 [astro-ph.HE]}}.

\bibitem{Fiorillo:2024upk}
D.~F.~G. Fiorillo and E.~Vitagliano, ``{Self-Interacting Dark Sectors in
  Supernovae Can Behave as a Relativistic Fluid},''
  \href{http://dx.doi.org/10.1103/PhysRevLett.133.251004}{{\em Phys. Rev.
  Lett.} {\bfseries 133} no.~25, (2024) 251004},
  \href{http://arxiv.org/abs/2404.07714}{{\ttfamily arXiv:2404.07714
  [hep-ph]}}.

\bibitem{Pospelov:2020ktu}
M.~Pospelov and H.~Ramani, ``{Earth-bound millicharge relics},''
  \href{http://dx.doi.org/10.1103/PhysRevD.103.115031}{{\em Phys. Rev. D}
  {\bfseries 103} no.~11, (2021) 115031},
  \href{http://arxiv.org/abs/2012.03957}{{\ttfamily arXiv:2012.03957
  [hep-ph]}}.

\bibitem{Berlin:2023zpn}
A.~Berlin, H.~Liu, M.~Pospelov, and H.~Ramani, ``{Terrestrial density of
  strongly-coupled relics},''
  \href{http://dx.doi.org/10.1103/PhysRevD.109.075027}{{\em Phys. Rev. D}
  {\bfseries 109} no.~7, (2024) 075027},
  \href{http://arxiv.org/abs/2302.06619}{{\ttfamily arXiv:2302.06619
  [hep-ph]}}.

\bibitem{Bowman:2018yin}
J.~D. Bowman, A.~E.~E. Rogers, R.~A. Monsalve, T.~J. Mozdzen, and N.~Mahesh,
  ``{An absorption profile centred at 78 megahertz in the sky-averaged
  spectrum},'' \href{http://dx.doi.org/10.1038/nature25792}{{\em Nature}
  {\bfseries 555} no.~7694, (2018) 67--70},
  \href{http://arxiv.org/abs/1810.05912}{{\ttfamily arXiv:1810.05912
  [astro-ph.CO]}}.

\bibitem{Barkana:2018qrx}
R.~Barkana, N.~J. Outmezguine, D.~Redigolo, and T.~Volansky, ``{Strong
  constraints on light dark matter interpretation of the EDGES signal},''
  \href{http://dx.doi.org/10.1103/PhysRevD.98.103005}{{\em Phys. Rev. D}
  {\bfseries 98} no.~10, (2018) 103005},
  \href{http://arxiv.org/abs/1803.03091}{{\ttfamily arXiv:1803.03091
  [hep-ph]}}.

\bibitem{Kovetz:2018zan}
E.~D. Kovetz, V.~Poulin, V.~Gluscevic, K.~K. Boddy, R.~Barkana, and
  M.~Kamionkowski, ``{Tighter limits on dark matter explanations of the
  anomalous EDGES 21 cm signal},''
  \href{http://dx.doi.org/10.1103/PhysRevD.98.103529}{{\em Phys. Rev. D}
  {\bfseries 98} no.~10, (2018) 103529},
  \href{http://arxiv.org/abs/1807.11482}{{\ttfamily arXiv:1807.11482
  [astro-ph.CO]}}.

\bibitem{Liu:2019knx}
H.~Liu, N.~J. Outmezguine, D.~Redigolo, and T.~Volansky, ``{Reviving
  Millicharged Dark Matter for 21-cm Cosmology},''
  \href{http://dx.doi.org/10.1103/PhysRevD.100.123011}{{\em Phys. Rev. D}
  {\bfseries 100} no.~12, (2019) 123011},
  \href{http://arxiv.org/abs/1908.06986}{{\ttfamily arXiv:1908.06986
  [hep-ph]}}.

\bibitem{Berlin:2018bsc}
A.~Berlin, N.~Blinov, G.~Krnjaic, P.~Schuster, and N.~Toro, ``{Dark Matter,
  Millicharges, Axion and Scalar Particles, Gauge Bosons, and Other New Physics
  with LDMX},'' \href{http://dx.doi.org/10.1103/PhysRevD.99.075001}{{\em Phys.
  Rev. D} {\bfseries 99} no.~7, (2019) 075001},
  \href{http://arxiv.org/abs/1807.01730}{{\ttfamily arXiv:1807.01730
  [hep-ph]}}.

\bibitem{Prinz:1998ua}
A.~A. Prinz {\em et~al.}, ``{Search for millicharged particles at SLAC},''
  \href{http://dx.doi.org/10.1103/PhysRevLett.81.1175}{{\em Phys. Rev. Lett.}
  {\bfseries 81} (1998) 1175--1178},
  \href{http://arxiv.org/abs/hep-ex/9804008}{{\ttfamily arXiv:hep-ex/9804008}}.

\bibitem{Magill:2018tbb}
G.~Magill, R.~Plestid, M.~Pospelov, and Y.-D. Tsai, ``{Millicharged particles
  in neutrino experiments},''
  \href{http://dx.doi.org/10.1103/PhysRevLett.122.071801}{{\em Phys. Rev.
  Lett.} {\bfseries 122} no.~7, (2019) 071801},
  \href{http://arxiv.org/abs/1806.03310}{{\ttfamily arXiv:1806.03310
  [hep-ph]}}.

\bibitem{Kelly:2018brz}
K.~J. Kelly and Y.-D. Tsai, ``{Proton fixed-target scintillation experiment to
  search for millicharged dark matter},''
  \href{http://dx.doi.org/10.1103/PhysRevD.100.015043}{{\em Phys. Rev. D}
  {\bfseries 100} no.~1, (2019) 015043},
  \href{http://arxiv.org/abs/1812.03998}{{\ttfamily arXiv:1812.03998
  [hep-ph]}}.

\bibitem{Harnik:2019zee}
R.~Harnik, Z.~Liu, and O.~Palamara, ``{Millicharged Particles in Liquid Argon
  Neutrino Experiments},''
  \href{http://dx.doi.org/10.1007/JHEP07(2019)170}{{\em JHEP} {\bfseries 07}
  (2019) 170}, \href{http://arxiv.org/abs/1902.03246}{{\ttfamily
  arXiv:1902.03246 [hep-ph]}}.

\bibitem{ArgoNeuT:2019ckq}
{\bfseries ArgoNeuT} Collaboration, R.~Acciarri {\em et~al.}, ``{Improved
  Limits on Millicharged Particles Using the ArgoNeuT Experiment at
  Fermilab},'' \href{http://dx.doi.org/10.1103/PhysRevLett.124.131801}{{\em
  Phys. Rev. Lett.} {\bfseries 124} no.~13, (2020) 131801},
  \href{http://arxiv.org/abs/1911.07996}{{\ttfamily arXiv:1911.07996
  [hep-ex]}}.

\bibitem{Kling:2022ykt}
F.~Kling, J.-L. Kuo, S.~Trojanowski, and Y.-D. Tsai, ``{FLArE up dark sectors
  with EM form factors at the LHC forward physics facility},''
  \href{http://dx.doi.org/10.1016/j.nuclphysb.2023.116103}{{\em Nucl. Phys. B}
  {\bfseries 987} (2023) 116103},
  \href{http://arxiv.org/abs/2205.09137}{{\ttfamily arXiv:2205.09137
  [hep-ph]}}.

\bibitem{SENSEI:2023gie}
{\bfseries SENSEI} Collaboration, L.~Barak {\em et~al.}, ``{Search by the
  SENSEI Experiment for Millicharged Particles Produced in the NuMI Beam},''
  \href{http://dx.doi.org/10.1103/PhysRevLett.133.071801}{{\em Phys. Rev.
  Lett.} {\bfseries 133} no.~7, (2024) 071801},
  \href{http://arxiv.org/abs/2305.04964}{{\ttfamily arXiv:2305.04964
  [hep-ex]}}.

\bibitem{Tsai:2024wdh}
Y.-D. Tsai, I.~Hwang, R.~Schmitz, M.~Citron, K.~Gunthoti, J.~Steenis, H.~Jeong,
  H.~Moon, J.~H. Yoo, and M.~X. Liu, ``{LANSCE-mQ: Dedicated search for
  milli/fractionally charged particles at LANL},''
  \href{http://arxiv.org/abs/2407.07142}{{\ttfamily arXiv:2407.07142
  [hep-ph]}}.

\bibitem{Essig:2024dpa}
R.~Essig, P.~Li, Z.~Liu, M.~McDuffie, R.~Plestid, and H.~Xu, ``{Probing
  millicharged particles at an electron beam dump with ultralow-threshold
  sensors},'' \href{http://dx.doi.org/10.1007/JHEP04(2025)057}{{\em JHEP}
  {\bfseries 04} (2025) 057}, \href{http://arxiv.org/abs/2412.09652}{{\ttfamily
  arXiv:2412.09652 [hep-ph]}}.

\bibitem{Plestid:2020kdm}
R.~Plestid, V.~Takhistov, Y.-D. Tsai, T.~Bringmann, A.~Kusenko, and
  M.~Pospelov, ``{New Constraints on Millicharged Particles from Cosmic-ray
  Production},'' \href{http://dx.doi.org/10.1103/PhysRevD.102.115032}{{\em
  Phys. Rev. D} {\bfseries 102} (2020) 115032},
  \href{http://arxiv.org/abs/2002.11732}{{\ttfamily arXiv:2002.11732
  [hep-ph]}}.

\bibitem{Harnik:2020ugb}
R.~Harnik, R.~Plestid, M.~Pospelov, and H.~Ramani, ``{Millicharged cosmic rays
  and low recoil detectors},''
  \href{http://dx.doi.org/10.1103/PhysRevD.103.075029}{{\em Phys. Rev. D}
  {\bfseries 103} no.~7, (2021) 075029},
  \href{http://arxiv.org/abs/2010.11190}{{\ttfamily arXiv:2010.11190
  [hep-ph]}}.

\bibitem{Du:2022hms}
M.~Du, R.~Fang, and Z.~Liu, ``{Millicharged particles from proton
  bremsstrahlung in the atmosphere},''
  \href{http://dx.doi.org/10.1007/JHEP08(2024)174}{{\em JHEP} {\bfseries 08}
  (2024) 174}, \href{http://arxiv.org/abs/2211.11469}{{\ttfamily
  arXiv:2211.11469 [hep-ph]}}.

\bibitem{Wu:2024iqm}
H.~Wu, E.~Hardy, and N.~Song, ``{Searching for heavy millicharged particles
  from the atmosphere},''
  \href{http://dx.doi.org/10.1103/PhysRevD.110.115037}{{\em Phys. Rev. D}
  {\bfseries 110} no.~11, (2024) 115037},
  \href{http://arxiv.org/abs/2406.01668}{{\ttfamily arXiv:2406.01668
  [hep-ph]}}.

\bibitem{TEXONO:2018nir}
{\bfseries TEXONO} Collaboration, L.~Singh {\em et~al.}, ``{Constraints on
  millicharged particles with low threshold germanium detectors at Kuo-Sheng
  Reactor Neutrino Laboratory},''
  \href{http://dx.doi.org/10.1103/PhysRevD.99.032009}{{\em Phys. Rev. D}
  {\bfseries 99} no.~3, (2019) 032009},
  \href{http://arxiv.org/abs/1808.02719}{{\ttfamily arXiv:1808.02719
  [hep-ph]}}.

\bibitem{CONNIE:2024off}
{\bfseries CONNIE, Atucha-II} Collaboration, A.~A. Aguilar-Arevalo {\em
  et~al.}, ``{Search for reactor-produced millicharged particles with
  Skipper-CCDs at the CONNIE and Atucha-II experiments},''
  \href{http://arxiv.org/abs/2405.16316}{{\ttfamily arXiv:2405.16316
  [hep-ex]}}.

\bibitem{TEXONO:2006spf}
{\bfseries TEXONO} Collaboration, H.~M. Chang {\em et~al.}, ``{Search of axions
  at the Kuo-Sheng nuclear power station with a high-purity germanium
  detector},'' \href{http://dx.doi.org/10.1103/PhysRevD.75.052004}{{\em Phys.
  Rev. D} {\bfseries 75} (2007) 052004},
  \href{http://arxiv.org/abs/hep-ex/0609001}{{\ttfamily arXiv:hep-ex/0609001}}.

\bibitem{Dent:2019ueq}
J.~B. Dent, B.~Dutta, D.~Kim, S.~Liao, R.~Mahapatra, K.~Sinha, and A.~Thompson,
  ``{New Directions for Axion Searches via Scattering at Reactor Neutrino
  Experiments},'' \href{http://dx.doi.org/10.1103/PhysRevLett.124.211804}{{\em
  Phys. Rev. Lett.} {\bfseries 124} no.~21, (2020) 211804},
  \href{http://arxiv.org/abs/1912.05733}{{\ttfamily arXiv:1912.05733
  [hep-ph]}}.

\bibitem{AristizabalSierra:2020rom}
D.~Aristizabal~Sierra, V.~De~Romeri, L.~J. Flores, and D.~K. Papoulias,
  ``{Axionlike particles searches in reactor experiments},''
  \href{http://dx.doi.org/10.1007/JHEP03(2021)294}{{\em JHEP} {\bfseries 03}
  (2021) 294}, \href{http://arxiv.org/abs/2010.15712}{{\ttfamily
  arXiv:2010.15712 [hep-ph]}}.

\bibitem{Kroll:1955zu}
N.~M. Kroll and W.~Wada, ``{Internal pair production associated with the
  emission of high-energy gamma rays},''
  \href{http://dx.doi.org/10.1103/PhysRev.98.1355}{{\em Phys. Rev.} {\bfseries
  98} (1955) 1355--1359}.

\bibitem{Pitrou:2019pqh}
C.~Pitrou and M.~Pospelov, ``{QED corrections to Big-Bang nucleosynthesis
  reaction rates},'' \href{http://dx.doi.org/10.1103/PhysRevC.102.015803}{{\em
  Phys. Rev. C} {\bfseries 102} no.~1, (2020) 015803},
  \href{http://arxiv.org/abs/1904.07795}{{\ttfamily arXiv:1904.07795
  [astro-ph.CO]}}.

\bibitem{talou2023nuclear}
P.~Talou and R.~Vogt, {\em Nuclear Fission: Theories, Experiments and
  Applications}.
\newblock Springer Nature, 2023.

\bibitem{TEXONO:2005fmk}
{\bfseries TEXONO} Collaboration, B.~Xin {\em et~al.}, ``{Production of
  electron neutrinos at nuclear power reactors and the prospects for neutrino
  physics},'' \href{http://dx.doi.org/10.1103/PhysRevD.72.012006}{{\em Phys.
  Rev. D} {\bfseries 72} (2005) 012006},
  \href{http://arxiv.org/abs/hep-ex/0502001}{{\ttfamily arXiv:hep-ex/0502001}}.

\bibitem{ENSDF}
 From ENSDF database as of December 14, 2024. Version available at
  http://www.nndc.bnl.gov/ensarchivals/.

\bibitem{LZ:2024iwc}
{\bfseries LZ} Collaboration, J.~Aalbers {\em et~al.}, ``{First search for
  atmospheric millicharged particles with the LUX-ZEPLIN experiment},''
  \href{http://arxiv.org/abs/2412.04854}{{\ttfamily arXiv:2412.04854
  [hep-ex]}}.

\bibitem{Allison:1980vw}
W.~W.~M. Allison and J.~H. Cobb, ``{Relativistic Charged Particle
  Identification by Energy Loss},''
  \href{http://dx.doi.org/10.1146/annurev.ns.30.120180.001345}{{\em Ann. Rev.
  Nucl. Part. Sci.} {\bfseries 30} (1980) 253--298}.

\bibitem{Henke:1993eda}
B.~L. Henke, E.~M. Gullikson, and J.~C. Davis, ``{X-Ray Interactions:
  Photoabsorption, Scattering, Transmission, and Reflection at E = 50-30,000
  eV, Z = 1-92},'' \href{http://dx.doi.org/10.1006/adnd.1993.1013}{{\em Atom.
  Data Nucl. Data Tabl.} {\bfseries 54} no.~2, (1993) 181--342}.

\bibitem{MUNU:2005xnz}
{\bfseries MUNU} Collaboration, Z.~Daraktchieva {\em et~al.}, ``{Final results
  on the neutrino magnetic moment from the MUNU experiment},''
  \href{http://dx.doi.org/10.1016/j.physletb.2005.04.030}{{\em Phys. Lett. B}
  {\bfseries 615} (2005) 153--159},
  \href{http://arxiv.org/abs/hep-ex/0502037}{{\ttfamily arXiv:hep-ex/0502037}}.

\bibitem{TEXONO:2009knm}
{\bfseries TEXONO} Collaboration, M.~Deniz {\em et~al.}, ``{Measurement of
  Nu(e)-bar -Electron Scattering Cross-Section with a CsI(Tl) Scintillating
  Crystal Array at the Kuo-Sheng Nuclear Power Reactor},''
  \href{http://dx.doi.org/10.1103/PhysRevD.81.072001}{{\em Phys. Rev. D}
  {\bfseries 81} (2010) 072001},
  \href{http://arxiv.org/abs/0911.1597}{{\ttfamily arXiv:0911.1597 [hep-ex]}}.

\bibitem{Ackermann:2025obx}
N.~Ackermann {\em et~al.}, ``{First observation of reactor antineutrinos by
  coherent scattering},'' \href{http://arxiv.org/abs/2501.05206}{{\ttfamily
  arXiv:2501.05206 [hep-ex]}}.

\bibitem{Borexino:2017rsf}
{\bfseries Borexino} Collaboration, M.~Agostini {\em et~al.}, ``{First
  Simultaneous Precision Spectroscopy of $pp$, $^7$Be, and $pep$ Solar
  Neutrinos with Borexino Phase-II},''
  \href{http://dx.doi.org/10.1103/PhysRevD.100.082004}{{\em Phys. Rev. D}
  {\bfseries 100} no.~8, (2019) 082004},
  \href{http://arxiv.org/abs/1707.09279}{{\ttfamily arXiv:1707.09279
  [hep-ex]}}.

\bibitem{XENON:2022ltv}
{\bfseries XENON} Collaboration, E.~Aprile {\em et~al.}, ``{Search for New
  Physics in Electronic Recoil Data from XENONnT},''
  \href{http://dx.doi.org/10.1103/PhysRevLett.129.161805}{{\em Phys. Rev.
  Lett.} {\bfseries 129} no.~16, (2022) 161805},
  \href{http://arxiv.org/abs/2207.11330}{{\ttfamily arXiv:2207.11330
  [hep-ex]}}.

\bibitem{LZ:2023poo}
{\bfseries LZ} Collaboration, J.~Aalbers {\em et~al.}, ``{Search for new
  physics in low-energy electron recoils from the first LZ exposure},''
  \href{http://dx.doi.org/10.1103/PhysRevD.108.072006}{{\em Phys. Rev. D}
  {\bfseries 108} no.~7, (2023) 072006},
  \href{http://arxiv.org/abs/2307.15753}{{\ttfamily arXiv:2307.15753
  [hep-ex]}}.

\bibitem{PandaX:2024cic}
{\bfseries PandaX} Collaboration, X.~Zeng {\em et~al.}, ``{Exploring New
  Physics with PandaX-4T Low Energy Electronic Recoil Data},''
  \href{http://dx.doi.org/10.1103/PhysRevLett.134.041001}{{\em Phys. Rev.
  Lett.} {\bfseries 134} no.~4, (2025) 041001},
  \href{http://arxiv.org/abs/2408.07641}{{\ttfamily arXiv:2408.07641
  [hep-ex]}}.

\bibitem{huang2013reference}
Y.~Huang, V.~Chubakov, F.~Mantovani, R.~L. Rudnick, and W.~F. McDonough, ``A
  reference earth model for the heat-producing elements and associated
  geoneutrino flux,'' {\em Geochemistry, Geophysics, Geosystems} {\bfseries 14}
  no.~6, (2013) 2003--2029.

\bibitem{Borexino:2015ucj}
{\bfseries Borexino} Collaboration, M.~Agostini {\em et~al.}, ``{Spectroscopy
  of geoneutrinos from 2056 days of Borexino data},''
  \href{http://dx.doi.org/10.1103/PhysRevD.92.031101}{{\em Phys. Rev. D}
  {\bfseries 92} no.~3, (2015) 031101},
  \href{http://arxiv.org/abs/1506.04610}{{\ttfamily arXiv:1506.04610
  [hep-ex]}}.

\bibitem{ParticleDataGroup:2024cfk}
{\bfseries Particle Data Group} Collaboration, S.~Navas {\em et~al.}, ``{Review
  of particle physics},''
  \href{http://dx.doi.org/10.1103/PhysRevD.110.030001}{{\em Phys. Rev. D}
  {\bfseries 110} no.~3, (2024) 030001}.

\bibitem{Ajzenberg-Selove:1991rsl}
F.~Ajzenberg-Selove, ``{Energy levels of light nuclei A = 13-15},''
  \href{http://dx.doi.org/10.1016/0375-9474(91)90446-D}{{\em Nucl. Phys. A}
  {\bfseries 523} (1991) 1--196}.

\bibitem{Marcucci:2015yla}
L.~E. Marcucci, G.~Mangano, A.~Kievsky, and M.~Viviani, ``{Implication of the
  proton-deuteron radiative capture for Big Bang Nucleosynthesis},''
  \href{http://dx.doi.org/10.1103/PhysRevLett.116.102501}{{\em Phys. Rev.
  Lett.} {\bfseries 116} no.~10, (2016) 102501},
  \href{http://arxiv.org/abs/1510.07877}{{\ttfamily arXiv:1510.07877
  [nucl-th]}}. [Erratum: Phys.Rev.Lett. 117, 049901 (2016)].

\bibitem{DEramo:2023buu}
F.~D'Eramo, G.~Lucente, N.~Nath, and S.~Yun, ``{Terrestrial detection of hidden
  vectors produced by solar nuclear reactions},''
  \href{http://dx.doi.org/10.1007/JHEP12(2023)091}{{\em JHEP} {\bfseries 12}
  (2023) 091}, \href{http://arxiv.org/abs/2305.14420}{{\ttfamily
  arXiv:2305.14420 [hep-ph]}}.

\bibitem{Oscura:2023qch}
{\bfseries Oscura} Collaboration, S.~Perez {\em et~al.}, ``{Searching for
  millicharged particles with 1 kg of Skipper-CCDs using the NuMI beam at
  Fermilab},'' \href{http://dx.doi.org/10.1007/JHEP02(2024)072}{{\em JHEP}
  {\bfseries 02} (2024) 072}, \href{http://arxiv.org/abs/2304.08625}{{\ttfamily
  arXiv:2304.08625 [hep-ex]}}.

\bibitem{DAMIC-M:2025luv}
{\bfseries DAMIC-M} Collaboration, K.~Aggarwal {\em et~al.}, ``{Probing
  Benchmark Models of Hidden-Sector Dark Matter with DAMIC-M},''
  \href{http://arxiv.org/abs/2503.14617}{{\ttfamily arXiv:2503.14617
  [hep-ex]}}.

\bibitem{Hagner:1995bn}
C.~Hagner, M.~Altmann, F.~von Feilitzsch, L.~Oberauer, Y.~Declais, and
  E.~Kajfasz, ``{Experimental search for the neutrino decay neutrino (3)
  ---\ensuremath{>} j-neutrino + e+ + e- and limits on neutrino mixing},''
  \href{http://dx.doi.org/10.1103/PhysRevD.52.1343}{{\em Phys. Rev. D}
  {\bfseries 52} (1995) 1343--1352}.

\bibitem{Bjorken:1988as}
J.~D. Bjorken, S.~Ecklund, W.~R. Nelson, A.~Abashian, C.~Church, B.~Lu, L.~W.
  Mo, T.~A. Nunamaker, and P.~Rassmann, ``{Search for Neutral Metastable
  Penetrating Particles Produced in the SLAC Beam Dump},''
  \href{http://dx.doi.org/10.1103/PhysRevD.38.3375}{{\em Phys. Rev. D}
  {\bfseries 38} (1988) 3375}.

\bibitem{Andreas:2012mt}
S.~Andreas, C.~Niebuhr, and A.~Ringwald, ``{New Limits on Hidden Photons from
  Past Electron Beam Dumps},''
  \href{http://dx.doi.org/10.1103/PhysRevD.86.095019}{{\em Phys. Rev. D}
  {\bfseries 86} (2012) 095019},
  \href{http://arxiv.org/abs/1209.6083}{{\ttfamily arXiv:1209.6083 [hep-ph]}}.

\bibitem{Marsicano:2018krp}
L.~Marsicano, M.~Battaglieri, M.~Bondi', C.~D.~R. Carvajal, A.~Celentano,
  M.~De~Napoli, R.~De~Vita, E.~Nardi, M.~Raggi, and P.~Valente, ``{Dark photon
  production through positron annihilation in beam-dump experiments},''
  \href{http://dx.doi.org/10.1103/PhysRevD.98.015031}{{\em Phys. Rev. D}
  {\bfseries 98} no.~1, (2018) 015031},
  \href{http://arxiv.org/abs/1802.03794}{{\ttfamily arXiv:1802.03794
  [hep-ex]}}.

\bibitem{Chang:2016ntp}
J.~H. Chang, R.~Essig, and S.~D. McDermott, ``{Revisiting Supernova 1987A
  Constraints on Dark Photons},''
  \href{http://dx.doi.org/10.1007/JHEP01(2017)107}{{\em JHEP} {\bfseries 01}
  (2017) 107}, \href{http://arxiv.org/abs/1611.03864}{{\ttfamily
  arXiv:1611.03864 [hep-ph]}}.

\bibitem{KamLAND:2011fld}
{\bfseries KamLAND} Collaboration, S.~Abe {\em et~al.}, ``{Measurement of the
  8B Solar Neutrino Flux with the KamLAND Liquid Scintillator Detector},''
  \href{http://dx.doi.org/10.1103/PhysRevC.84.035804}{{\em Phys. Rev. C}
  {\bfseries 84} (2011) 035804},
  \href{http://arxiv.org/abs/1106.0861}{{\ttfamily arXiv:1106.0861 [hep-ex]}}.

\bibitem{Firestone:2017fet}
R.~B. Firestone, T.~Belgya, M.~Krti{\v{c}}ka, F.~Be{\v{c}}v{\'a}{\v{r}},
  L.~Szentmiklo{\textperiodcentered}si, and I.~Tomandl, ``{Thermal neutron
  capture cross section for Fe56(n,{\ensuremath{\gamma}})},''
  \href{http://dx.doi.org/10.1103/PhysRevC.95.014328}{{\em Phys. Rev. C}
  {\bfseries 95} no.~1, (2017) 014328}.

\bibitem{Waites:2022tov}
L.~Waites, A.~Thompson, A.~Bungau, J.~M. Conrad, B.~Dutta, W.-C. Huang, D.~Kim,
  M.~Shaevitz, and J.~Spitz, ``{Axionlike particle production at beam dump
  experiments with distinct nuclear excitation lines},''
  \href{http://dx.doi.org/10.1103/PhysRevD.107.095010}{{\em Phys. Rev. D}
  {\bfseries 107} no.~9, (2023) 095010},
  \href{http://arxiv.org/abs/2207.13659}{{\ttfamily arXiv:2207.13659
  [hep-ph]}}.

\bibitem{JUNO:2015zny}
{\bfseries JUNO} Collaboration, F.~An {\em et~al.}, ``{Neutrino Physics with
  JUNO},'' \href{http://dx.doi.org/10.1088/0954-3899/43/3/030401}{{\em J. Phys.
  G} {\bfseries 43} no.~3, (2016) 030401},
  \href{http://arxiv.org/abs/1507.05613}{{\ttfamily arXiv:1507.05613
  [physics.ins-det]}}.

\bibitem{Pospelov:2008jk}
M.~Pospelov, A.~Ritz, and M.~B. Voloshin, ``{Bosonic super-WIMPs as keV-scale
  dark matter},'' \href{http://dx.doi.org/10.1103/PhysRevD.78.115012}{{\em
  Phys. Rev. D} {\bfseries 78} (2008) 115012},
  \href{http://arxiv.org/abs/0807.3279}{{\ttfamily arXiv:0807.3279 [hep-ph]}}.

\bibitem{McDermott:2017qcg}
S.~D. McDermott, H.~H. Patel, and H.~Ramani, ``{Dark Photon Decay Beyond The
  Euler-Heisenberg Limit},''
  \href{http://dx.doi.org/10.1103/PhysRevD.97.073005}{{\em Phys. Rev. D}
  {\bfseries 97} no.~7, (2018) 073005},
  \href{http://arxiv.org/abs/1705.00619}{{\ttfamily arXiv:1705.00619
  [hep-ph]}}.

\bibitem{Pospelov:2017kep}
M.~Pospelov and Y.-D. Tsai, ``{Light scalars and dark photons in Borexino and
  LSND experiments},''
  \href{http://dx.doi.org/10.1016/j.physletb.2018.08.053}{{\em Phys. Lett. B}
  {\bfseries 785} (2018) 288--295},
  \href{http://arxiv.org/abs/1706.00424}{{\ttfamily arXiv:1706.00424
  [hep-ph]}}.

\bibitem{Brdar:2020dpr}
V.~Brdar, B.~Dutta, W.~Jang, D.~Kim, I.~M. Shoemaker, Z.~Tabrizi, A.~Thompson,
  and J.~Yu, ``{Axionlike Particles at Future Neutrino Experiments: Closing the
  Cosmological Triangle},''
  \href{http://dx.doi.org/10.1103/PhysRevLett.126.201801}{{\em Phys. Rev.
  Lett.} {\bfseries 126} no.~20, (2021) 201801},
  \href{http://arxiv.org/abs/2011.07054}{{\ttfamily arXiv:2011.07054
  [hep-ph]}}.

\bibitem{NEON:2024kwv}
{\bfseries NEON} Collaboration, B.~J. Park {\em et~al.}, ``{New Constraints on
  Axionlike Particles with the NEON Detector at a Nuclear Reactor},''
  \href{http://dx.doi.org/10.1103/PhysRevLett.134.201002}{{\em Phys. Rev.
  Lett.} {\bfseries 134} no.~20, (2025) 201002},
  \href{http://arxiv.org/abs/2406.06117}{{\ttfamily arXiv:2406.06117
  [hep-ex]}}.

\bibitem{Danilov:2018bks}
M.~Danilov, S.~Demidov, and D.~Gorbunov, ``{Constraints on hidden photons
  produced in nuclear reactors},''
  \href{http://dx.doi.org/10.1103/PhysRevLett.122.041801}{{\em Phys. Rev.
  Lett.} {\bfseries 122} no.~4, (2019) 041801},
  \href{http://arxiv.org/abs/1804.10777}{{\ttfamily arXiv:1804.10777
  [hep-ph]}}.

\end{thebibliography}\endgroup

\end{document}